\documentclass[10pt]{article}
\usepackage{amsmath}
\usepackage{amsthm}
\usepackage{amssymb}
\usepackage{subfigure}
\newtheorem{definition}{Definition}

\newtheorem{theorem}[definition]{Theorem}
\newtheorem{corollary}[definition]{Corollary}

\newtheorem{lemma}[definition]{Lemma}
\newtheorem{claim}[definition]{Claim}
\newtheorem{fact}[definition]{Fact}

\usepackage{typearea}
\newcommand{\SD}[2]{\|#1 - #2\|}
\newcommand*{\n}{\penalty 100\relax} 
\newcommand{\eps}{\varepsilon}

\newcommand*{\Pd}{\mathsf{P}}

\DeclareMathOperator*{\E}{E}

\newcommand*{\vbar}{\bar{v}}
\newcommand*{\xbar}{\bar{x}}
\newcommand*{\ybar}{\bar{y}}

\newcommand*{\Ubar}{\overline{U}}
\newcommand*{\Sbar}{\overline{S}}
\newcommand*{\Tbar}{\overline{T}}

\newcommand*{\Xbar}{\overline{X}}
\newcommand*{\Ybar}{\overline{Y}}
\newcommand*{\cA}{\mathcal{A}}
\newcommand*{\cB}{\mathcal{B}}

\newcommand*{\cR}{\mathcal{R}}
\newcommand*{\cS}{\mathcal{S}}
\newcommand*{\cT}{\mathcal{T}}
\newcommand*{\cU}{\mathcal{U}}
\newcommand*{\cV}{\mathcal{V}}
\newcommand*{\cX}{\mathcal{X}}
\newcommand*{\cY}{\mathcal{Y}}
\newcommand*{\cZ}{\mathcal{Z}}
\newcommand*{\Atilde}{\widetilde{A}}
\newcommand*{\Btilde}{\widetilde{B}}
\newcommand*{\Stilde}{\widetilde{S}}
\newcommand*{\Ttilde}{\widetilde{T}}
\newcommand*{\Utilde}{\widetilde{U}}
\newcommand*{\Vtilde}{\widetilde{V}}
\newcommand*{\Xtilde}{\widetilde{X}}
\newcommand*{\Ytilde}{\widetilde{Y}}

\newcommand{\mkG}{\mathfrak{G}}

\newcommand{\Wevent}{W}
\newcommand{\NoJ}{{(\backslash j)}}
\newcommand{\Tot}{\mathrm{Tot}}

\newcommand{\vns}{v_{ns}}

\newcommand{\cancel}[1]{}

\title{Parallel Repetition: \\
  Simplifications and the No-Signaling Case}
\author{Thomas Holenstein\thanks{Microsoft Research, Silicon Valley; 
      \texttt{thomahol@microsoft.com}. This work was done while the author was at ETH Zurich.}}
\date{}
\begin{document}
\maketitle
\begin{abstract}
  Consider a game where a referee chooses~$(x,y)$ according
  to a publicly known distribution~$\Pd_{XY}$, sends~$x$ to Alice, and
  $y$ to Bob.  Without communicating with each other, Alice responds
  with a value~$a$ and Bob responds with a value~$b$.  Alice and Bob
  jointly win if a publicly known predicate~$Q(x,y,a,b)$ holds.
  
  Let such a game be given and assume that the maximum probability
  that Alice and Bob can win is~$v<1$.  Raz (SIAM J.~Comput.~27, 1998)
  shows that if the game is repeated~$n$ times in parallel, then the
  probability that Alice and Bob win \emph{all} games simultaneously
  is at most~$\vbar^{\tfrac{n}{\log(s)}}$, where $s$ is the maximal
  number of possible responses from Alice and Bob in the initial game,
  and $\vbar<1$ is a constant depending only on~$v$.  
  
  In this work, we simplify Raz's proof in various ways and thus
  shorten it significantly.  Further we study the case where Alice and
  Bob are not restricted to local computations and can use any
  strategy which does not imply communication among them.
\end{abstract}
\section{Introduction}


The question how much parallel repetition of a game as in the abstract
reduces the winning probability of the players was motivated by the
study of two-prover interactive proofs, initiated by Ben-Or et al.
\cite{BGKW88}.  It was first conjectured that in a game which is
repeated~$n$ times in parallel, the probability that Alice and Bob win
all the games simultaneously is at most~$v^n$ (see \cite{FoRoSi94}).
However, later a counterexample to this conjecture was given
\cite{Fortno89}.

\paragraph{Related Work}
Various papers give upper bounds on the winning probability of a game
which is repeated~$n$ times in parallel
\cite{CaCoLi92,Feige91,LapSha95,Raz98,Verbit94}.  However, the upper
bound given by Raz \cite{Raz98} is the only explicit bound for
arbitrary distributions~$\Pd_{XY}$ (it is also quantitatively the
strongest).  Parnafes, Raz, and Wigderson \cite{PaRaWi97} modify Raz's
proof to show that the term~$\log(s)$ can be replaced by a parameter
which is much smaller for some games.

Games for which the~$n$-fold parallel repetition decreases the winning
probability less than from~$v$ to~$v^n$ were also constructed: Fortnow
\cite{Fortno89} gives\footnote{For readers not familiar with such
  counter-examples, a variation of Fortnow's game is reproduced in
  Appendix~\ref{app:nontriviality}.} a game for which the maximal
winning probability in two repetitions is larger than~$v^2$ (see also
\cite{FeiLov92}), Feige \cite{Feige91} constructs a game where the
winning probability in two parallel repetitions does not decrease at
all, and Feige and Verbitsky \cite{FeiVer02} give, for infinitely
many~$s$, a game where~$\Theta(\frac{\log(s)}{\log{\log(s)}})$
repetitions decrease the winning probability from at most~$\frac34$ to
at least~$\frac18$, where $s$ is the number of possible answers Alice
and Bob can give.  This last result shows that in general Raz's bound
is close to optimal.

\paragraph{No-signaling strategies}
No-signaling strategies are all those strategies which do not imply
communication.  Popescu and Rohrlich \cite{PopRoh94} give an example
of such a strategy: Alice receives a bit~$x$, Bob receives a bit~$y$,
and they respond with uniform random bits~$a$ and~$b$ such
that~$a\oplus b = x\land y$.  Note that even though we cannot
implement this strategy with shared randomness and without
communication, Alice and Bob cannot communicate if they only have
black-box access to such functionality.

The study of no-signaling strategies is motivated by the idea that if
Alice and Bob share some entangled quantum state, the set of possible
strategies they might use increases, but stays a subset of the
no-signaling strategies (this subset is strict: for example the above
strategy which achieves~$a\oplus b=x\land y$ from~$(x,y)$ cannot be
simulated perfectly using quantum mechanics \cite[Problem
2.3]{NieChu00}, \cite{Cirels80} --- the corresponding game is called
the CHSH-game \cite{CHSH69}).  

We remark that there are games which can be won with probability 1
given a shared quantum state (and thus with a no-signaling strategy),
but not using local strategies.  Those are called ``pseudo-telepathy
games'' (see \cite{BrBrTa05} and the references therein).

A parallel repetition theorem for the case where Alice and Bob share
a quantum state and the decision of the referee only depends on the
XOR of the binary answers of Alice and Bob was recently given by
Cleve et al.~\cite{CSUU06}.

\paragraph{Contributions of this paper}
In this paper we simplify Raz's proof.  Most importantly, we replace a
large part (essentially Section 6) of Raz's paper with the simpler
Lemma~\ref{lem:LocallyComputableCommonPartPre}.  This also allows us to
give an explicit bound on the maximal winning probability of a game
repeated~$n$ times in parallel (Raz does not explicitly describe the
dependence of~$\vbar$ on~$v$).  

The use of Lemma~\ref{lem:LocallyComputableCommonPartPre} also makes
the rest of the argument simpler.  We shortly explain why:  The main
part of the proof consists of showing that the information the players
get in the~$n$-fold repetition does not help them to win the subgame
in some coordinate~$j$, even conditioned on the event that certain
other subgames are won.  This is done in three steps.  In two of these
steps the information does not help the players because they can
generate this information themselves with local computation only.
Lemma~\ref{lem:LocallyComputableCommonPartPre} shows that this also holds
for the third step.  This allows us to merge some of the steps, which
simplifies the overal structure.

We also study how much the term~$\log(s)$ in the exponent in the
parallel repetition theorem can be reduced.  In \cite{PaRaWi97} it is
shown that the logarithm of the partition number of the accepance
predicate can be used instead of~$\log(s)$.  Based on the ideas from
there, Theorem~\ref{thm:pr_local_strengthened} gives a bound which
might be stronger for some games.

Finally, we prove a parallel repetition theorem in case Alice and Bob
are restricted to no-signaling strategies (in both the given game and
the parallel repetition of it).

\section{Notation and Basic Facts}
\subsection{Probability Distributions}\label{sec:probdistr}

We use calligraphic letters to denote sets.  We denote random
variables using capital letters, and values with lower case letters.
We use superscripts to denote tuples, e.g., $X^n := (X_1, \ldots,
X_{n})$ and $x^n :=\n (x_1, \ldots, x_{n})$.

If a distribution $\Pd_{XY}$ over~$\cX \times \cY$ is given, we write
$\Pd_X$ or~$\Pd_Y$ to denote the marginal distribution, e.g.,
$\Pd_X(x) := \n\sum_{y\in\cY}\Pd_{XY}(x,y)$.  The conditional
distribution~$\Pd_{Y|X=x}$ is $\Pd_{Y|X=x}(y) :=
\Pd_{XY}(x,y)/\Pd_X(x)$.

Let~$\Pd_{X_0}$ be a distribution over~$\cX$ and~$\Pd_{Y_1|X_1=x}$ be
a conditional distribution over~$\cY$.  We define the
distribution~$\Pd_{X_0}\Pd_{Y_1|X_1}$ over~$\cX\times\cY$ as
\begin{align}\label{eq:29}
  (\Pd_{X_0} \Pd_{Y_1|X_1})(x,y) := \Pd_{X_0}(x) \cdot \Pd_{Y_1|X_1=x}(y).
\end{align}
For this, it is necessary that $\Pd_{Y_1|X_1=x}$ is defined for
every~$x\in\cX$.  We also use this notation when~$\Pd_{Y_1|X_1=x}$ is
defined as marginal of a given distribution $\Pd_{X_1Y_1}$.  In this
case, we define~$\Pd_{Y_1|X_1=x}$ in an arbitrary way
if~$\Pd_{X_1}(x)=0$.  This notation is used for example in
Corollary~\ref{cor:LocallyComputableCommonPart} in the form
$\Pd_{X_0Y_0}\Pd_{S|X}$, where it is understood as
$(\Pd_{X_0Y_0}\Pd_{S|X})(x,y,s) := \Pd_{X_0Y_0}(x,y)\Pd_{S|X=x}(s)$.
Note that the conditional distribution~$\Pd_{S|X=x}$ is defined
there by the marginal distribution~$\Pd_{SX}$ of the given
distribution~$\Pd_{SXY}$.  Our notation is not explicit since it does
not specify which random variables are associated with each other.
However, this will always be clear from the context.

For two probability distributions~$\Pd_{X_0}$ and~$\Pd_{X_1}$
over the same set~$\cX$ we define the statistical distance
\begin{align}
  \|\Pd_{X_0} - \Pd_{X_1}\| := 
  \frac12\sum_{x\in\cX} \bigl|\Pd_{X_0}(x)-\Pd_{X_1}(x)\bigr|.
\end{align}
\subsection{Games}
\begin{definition}\label{def:game}
  A \emph{game~$\mkG=(\Pd_{XY},Q)$
    over~$\cX\times\cY\times\cA\times\cB$} is a
  distribution~$\Pd_{XY}$ over~$\cX \times\cY$ and a predicate~$Q$
  over~$\cX\times\cY\times\cA\times\cB$.  The \emph{value}~$v(\mkG)$
  of a game is
  \begin{align*}
    v(\mkG) := \max_{h_a,h_b} \Pr_{XY}[Q(X,Y,h_a(X),h_b(Y))],
  \end{align*}
  where the maximization is over functions~$h_a: \cX \rightarrow \cA$
  and~$h_b: \cY\rightarrow\cB$.  
  A \emph{strategy}~$(h_a,h_b)$ for a game is a pair of such functions.
\end{definition}

Sometimes also randomized strategies for Alice and Bob are considered,
where~$h_a$ and~$h_b$ also depend on (the same) shared randomness~$r$ chosen
according to some distribution~$\Pd_{R}$.  However, there always exists
an~$r\in\cR$ such that
\begin{align}
  \Pr_{RXY}[Q(X,Y,h_a(X,R),h_b(Y,R))]
  &=
  \E_R\bigl[\Pr_{XY}[Q(X,Y,h_a(X,R),h_b(Y,R))]\bigr]\nonumber\\
  &\leq
  \Pr_{XY}[Q(X,Y,h_a(X,r),h_b(Y,r))],\label{eq:5}
\end{align}
and we see that the definition of the value is robust against such a
change.  Individual (local) randomness can be obtained from shared
randomness and is thus a special case of the above.

\begin{definition}\label{def:nfoldrepetition}
  The~\emph{$n$-fold parallel repetition}~$\mkG^{n}$ of a game~$\mkG =
  (\Pd_{XY},Q)$ over~$\cX\times\cY\times\cA\times\cB$ is the game over
  $\cX^n\times\cY^n\times\cA^n\times\cB^n$ which is given by~$\mkG^{n}
  := (\Pd_{X^nY^n}, Q^{\land n})$ where
  \begin{align*}
    \Pd_{X^nY^n}(x^n,y^n) &:=
    \prod_{i=1}^n\Pd_{XY}(x_i,y_i),\text{\quad and}\\
    Q^{\land n}(x^n,y^n,a^n,b^n) &:= \bigwedge_{i=1}^n Q(x_i,y_i,a_i,b_i).
  \end{align*}
\end{definition}  
If a strategy is given, the distribution~$\Pd_{X^nY^nA^nB^n}$ of
queries and answers is defined in the obvious way.  We further define,
for all~$i$, the event~$W_i$ which occurs if the~$i$th subgame is won.
\begin{definition}\label{def:randomvars}
  For a game~$\mkG^{n}$ and a strategy~$(h_a,h_b)$ the
  distribution~$\Pd_{X^n Y^n A^n B^n}$
  over~$\cX^n\times\cY^n\times\cA^n \times\cB^n$ is given by
  \begin{align*}
    \Pd_{X^n Y^n A^n B^n}(x^n, y^n,a^n,b^n)
    &:=
    \begin{cases}
       \Pd_{X^nY^n}(x^n,y^n) & \text{if $h_a(x^n)=a^n$ and
        $h_b(y^n) = b^n$}\\
      0& \text{otherwise.}\\
    \end{cases}
  \end{align*}
  Further, $W^n$ is the tuple of events~$(W_1,\ldots,W_n)$ where~$W_i
  :\iff Q(X_i,Y_i,A_i,B_i)$.
\end{definition} 

We prove the following version of the parallel repetition theorem.
\begin{theorem}[Parallel Repetition Theorem]\label{thm:pr_local}
  For any game~$\mkG$ with value~$v := v(\mkG)$ and any integer~$n$:
  \begin{align*}
    v(\mkG^{ n}) \leq
    \Bigl(1-\frac{(1-v)^3}{6000}\Bigr)^{\frac{n}{\log(|\cA||\cB|)}}.
  \end{align*}
\end{theorem}
The constant 6000 could be improved by a more carful analysis (we will
not optimize constants which would improve it during the proof).
However, we do not know whether the 3 in the exponent can be reduced.

In \cite{PaRaWi97} it is shown that in Raz's proof the term
$\log(|\cA||\cB|)$ in the exponent can be reduced to the maximum of
the logarithm of the partition number of $Q(x,y,\cdot,\cdot)$.  As
shown by Beame \cite{Beame06}, the argument can be adapted to work
with the proof given here.  We give a slightly different argument in
Section~\ref{sec:reducingtheexponent} which shows how the term can be
reduced to a quantity which is a lower bound on the logarithm of the
partition number.

\section{Proof Sketch}
Fix an arbitrary game~$\mkG$, its $n$-fold parallel
repetition~$\mkG^{n}$, and a strategy~$h_a$,~$h_b$ for~$\mkG^{n}$.
With the notation from Definition~\ref{def:randomvars}, the parallel
repetition theorem is simply an upper bound on $\Pr[W_1\land\dots\land
W_n]$.  To get such an upper bound, we show that for arbitrary
indices~$i_1,\ldots,i_m$ there exists an index~$j$ such that
\begin{align}
  \Pr[W_{j}|W_{i_1}\land\dots\land W_{i_m}]\leq v(\mkG) + \eps, \label{eq:1}
\end{align}
where $\eps$ depends on~$m$,~$n$, $\log(|\cA||\cB|)$, and
$\Pr[W_{i_1}\land\dots\land W_{i_m}]$ (this is
Lemma~\ref{lem:gameconditionedvalue}).  From~(\ref{eq:1}) a simple
induction gives the parallel repetition theorem, thus we now
concentrate on the proof of~(\ref{eq:1}).

\paragraph{Locally Computable Embeddings}
In order to prove~(\ref{eq:1}) we define the
distribution
\begin{align}\label{eq:10}
  \Pd_{\Xtilde^n \Ytilde^n} := \Pd_{X^nY^n|W_{i_1}\land\dots\land
    W_{i_m}}
\end{align}
(i.e., the distribution of the message which the referee sends to
Alice and Bob conditioned on the event that the games~$i_1$ to~$i_m$
are won). 

We show (Lemma~\ref{lem:locallyComputableGame}) that for some~$j$ the
following can be achieved by Alice and Bob without communication and
using shared randomness only:
\begin{enumerate}
\item Alice, on input~$x$, produces a tuple~$\xbar^n$
  with~$\xbar_j = x$.
\item Bob, on input~$y$, produces a tuple~$\ybar^n$ with~$\ybar_j
  = y$.
\item Let $\Pd_{\Xbar^n\Ybar^n}$ be the resulting joint distribution
  of the tuples~$(\xbar^n,\ybar^n)$, assuming that $(x,y)$ is chosen
  according to $\Pd_{XY}$.  Then
  \begin{align*}
    \| \Pd_{\Xbar^n\Ybar^n} - \Pd_{\Xtilde^n\Ytilde^n} \| \leq \eps.
  \end{align*}
\end{enumerate}
We say that~$(X,Y)$ can be $1-\eps$-embedded into
$(\Xtilde^n,\Ytilde^n)$ with~$(\Xtilde_j,\Ytilde_j)=(X,Y)$ by local
computation.  

If such an embedding is given, we can consider the following strategy
for the initial game~$\mkG$: Alice and Bob embed their inputs $(X,Y)$
in $(\Xtilde^n,\Ytilde^n)$ with~$(\Xtilde_j,\Ytilde_j)=(X,Y)$, and
answer with coordinate~$j$ of~$h_a(\Xtilde^n)$ and~$h_b(\Ytilde^n)$.
This strategy wins with probability at
least~$\Pr[W_j|W_{i_1}\land\dots\land W_{i_m}]-\eps$.  Since no
strategy for the initial game has higher winning probability than
$v(\mkG)$ this implies (\ref{eq:1}).

We remark that a necessary condition for such an embedding to exist is
that
\begin{align}\label{eq:2}
  \| \Pd_{XY} - \Pd_{\Xtilde_j \Ytilde_j} \| \leq \eps,
\end{align}
and indeed this follows from Lemma~\ref{lem:basicrepetitions} for~$U_j
= (X_j,Y_j)$ (of course this condition is not a sufficient one).

\paragraph{Constructing an Embedding}
We now give a more detailled explanation how Alice and Bob can
embed~$(X,Y)$ into~$(\Xtilde^n,\Ytilde^n)$
with~$(\Xtilde_j,\Ytilde_j)=(X,Y)$.  For this, given values~$(x,y)$
distributed according to~$\Pd_{XY}$, Alice and Bob proceed as follows:
\begin{enumerate}
\item Alice and Bob use shared randomness to produce queries and
  responses for all the won games, i.e., 
  values~$(x_{i_1},y_{i_1},a_{i_1},b_{i_1})$
  to~$(x_{i_m},y_{i_m},a_{i_m},b_{i_m})$.  Here, Alice and
  Bob \emph{both} produce \emph{all} these values.
\item For every index~$i \notin \{i_1,\ldots,i_m,j\}$, Alice and Bob
  examine a shared random bit~$d_i$.  If~$d_i = 1$ both locally
  produce~$x_i$, otherwise both locally produce~$y_i$.  Again, Alice
  and Bob both produce all these values.
\item Using individual randomness, Alice and Bob locally expand their
  information such that Alice gets~$x^n$ and Bob~$y^n$.
\end{enumerate}
In steps 1 and 2 we have to take care of two things: first, the values
produced should be distributed according to the the respective
marginal of the distribution
$\Pd_{\Atilde^n\Btilde^n\Xtilde^n\Ytilde^n|\Xtilde_j=x \land
  \Ytilde_j=y}$ (where $\Pd_{\Atilde^n\Btilde^n\Xtilde^n\Ytilde^n}$ is
defined analogously to (\ref{eq:10})).  Second, Alice and Bob should
produce \emph{equal values} (otherwise the resulting random
variables~$(\Xbar^n,\Ybar^n)$ will not have the correct overall
distribution).

For step 1 achieving both is simple: it follows from
Corollary~\ref{cor:disjointreps} that Alice and Bob can choose the
values~$(x_{i_1},y_{i_1},a_{i_1},b_{i_1}),\ldots,(x_{i_m},y_{i_m},a_{i_m},b_{i_m})$
independently of~$(x,y)$ according to
$\Pd_{\Xtilde_{i_1}\Ytilde_{i_1}\Atilde_{i_1}\Btilde_{i_1} \cdots
  \Xtilde_{i_m}\Ytilde_{i_m}\Atilde_{i_m}\Btilde_{i_m}}$.  Using
shared randomness this can be done such that both get the same tuple.

The second step is harder, as in this case the values cannot be chosen
independently of~$(x_j,y_j)$ anymore.\footnote{The values~$d_i$ can be
  chosen independently, but \emph{not} the values of $x_i$ respective
  $y_i$.  We quickly explain why this is impossible in general.
  Assume that the random variables~$X$ and~$Y$ contain a shared
  bit~$B$.  The game~$\mkG^n$ and the strategy~$(h_a,h_b)$ may be such
  that Alice and Bob win subgame~$i_1$ in case~$B_1\oplus\dots\oplus
  B_n=0$.  Generating the values independently of~$(x,y)$ would now
  produce a distribution with statistical distance at least~$\frac12$
  from the target distribution.  Therefore, a bit which is contained
  in both $x$ and~$y$ \emph{must} be considered when generating the
  values of~$x_i$ and~$y_i$.}  However, let~$\Stilde$ be the random
variables which Alice and Bob produce in this step.  It will follow
from Corollary~\ref{cor:disjointreps} that
$\|\Pd_{XY}\Pd_{\Stilde|\Xtilde_j} - \Pd_{XY\Stilde}\|$ and
$\|\Pd_{XY}\Pd_{\Stilde|\Ytilde_j} - \Pd_{XY\Stilde}\|$ are both
small, and Lemma~\ref{lem:LocallyComputableCommonPartPre} implies that
this is sufficient to generate~$\Stilde$ locally.

In fact, Corollary~\ref{cor:disjointreps}
and Lemma~\ref{lem:LocallyComputableCommonPartPre} are strong enough to do
steps 1 and 2 at the same time, and thus these steps are done
simultaneously in the proof of Lemma~\ref{lem:locallyComputableGame}.

Step 3 will be simpler to implement.  Because the players also
computed~$a_{i_\ell}$ and~$b_{i_\ell}$ in step 1, they can expand
their known values according to the given distributions and the
resulting distribution will be correct (this follows from
Lemma~\ref{lem:locallyComputableMarkov}, and a detailed explanation
is in the proof of Lemma~\ref{lem:locallyComputableGame}).

\section{Conditioned Distributions}

The following lemma is essentially Claim~5.1 in Raz's paper
\cite{Raz98} (and we use the proof given there).  It states that if
random variables $U_i$ are chosen independently, then conditioning on
an event does not change the individual distributions a lot on
average.

\begin{lemma}\label{lem:basicrepetitions}
  Let $\Pd_{U^k} := \Pd_{U_1} \dots \Pd_{U_k}$ be a probability
  distribution over~$\cU^k$, $\Wevent$ an event.  Then,
  \begin{align}\label{eq:50}
    \Pr[W] \leq 2^{-\sum_{j=1}^k (\SD{\Pd_{U_j|\Wevent}}{\Pd_{U_j}})^2}.
  \end{align}
\end{lemma}
As an example, let $U_i$ be uniform and independent bits and~$W$ be
the event that at least~$k(\frac{1}{2}+\eps)$ of these bits are one.
Then $\|P_{U_i|W} - \Pd_{U_i}\| \geq \eps$ and the lemma states that
$\Pr[W] \leq 2^{-k\eps^2}$, which is a version of Chernoff's
inequality (note that this implies that
Lemma~\ref{lem:basicrepetitions} is almost tight; see, for example,
\cite{HolRen06}).

Using $(\sum_{j=1}^k a_j)^2 \leq k \sum_{j=1}^k a_j^2$ one easily checks that
\eqref{eq:50} implies
\begin{align}\label{eq:52}
  \sum_{j=1}^k \| \Pd_{U_j|W} - \Pd_{U_j}\| \leq 
  \sqrt{k\log\Bigl(\frac{1}{\Pr[W]}\Bigr)}\;,
\end{align}
which is the form we use later.

\begin{proof}
  For two distributions~$\Pd_{S}$ and~$\Pd_{T}$ over the same
  set~$\cS$, the \emph{relative entropy} $D(\Pd_{S}\|\Pd_{T})$ is
  defined as
  \begin{align}\label{eq:8}
    D(\Pd_{S}\|\Pd_{T}) := \sum_{s \in \cS} \Pd_{S}(s)\log\Bigl(\frac{\Pd_{S}(s)}{\Pd_{T}(s)}\Bigr).
  \end{align}
  This quantity satisfies $D(\Pd_{S}\|\Pd_{T}) \geq
  \bigl(\|\Pd_{S}-\Pd_{T}\|\bigr)^2$ (see \cite[Lemma
  12.6.1]{CovTho91}).  Also, if~$\Pd_{U^k} = \Pd_{U_1}\dots
  \Pd_{U_k}$ and~$\Pd_{V^k}$ are distributions over the set~$\cU^k$,
  then $\sum_{j=1}^k D(\Pd_{V_j}\|\Pd_{U_j}) \leq
  D(\Pd_{V^k}\|\Pd_{U^k})$ (see Appendix~\ref{app:relentrlemma}).
  
  Using the above we get
  \begin{align*}
    \sum_{j=1}^k\Bigl(\SD{\Pd_{U_j|\Wevent}}{\Pd_{U_j}}\Bigr)^2
    &\leq \sum_{j=1}^k D(\Pd_{U_j|\Wevent}\|{\Pd_{U_j}})\\
    &\leq D(\Pd_{U^k|\Wevent}\|\Pd_{U^k})\displaybreak[2]\\
    &= \sum_{u^k} \Pd_{U^k|\Wevent}(u^k)
    \log\Bigl(\frac{\Pd_{U^k|\Wevent}(u^k)}{\Pd_{U^k}(u^k)}\Bigr)\\
    &= \sum_{u^k} \Pd_{U^k|\Wevent}(u^k)
    \log\Bigl(\frac{\Pr[\Wevent|U^k=u^k]}{\Pr[\Wevent]}\Bigr)\\
    &= \log\Bigl(\frac{1}{\Pr[\Wevent]}\Bigr) +
     \sum_{u^k} \Pd_{U^k|\Wevent}(u^k)
    \log\bigl(\Pr[\Wevent|U^k=u^k]\bigr)\\
    &\leq \log\Bigl(\frac{1}{\Pr[\Wevent]}\Bigr).\qedhere
  \end{align*}
\end{proof}

We now give a slight extension of this lemma (this makes it simpler to
apply later).  First, the~$U_j$ are independent given the value of an
additional random variable~$T$.  Second, an arbitrary third random
variable~$V$ with bounded alphabet size gives side information
about~$U_j$.  Then, choosing~$U_j$ without considering the fact that
an event~$W$ happened and ignoring~$V$ does not change the
distribution of~$U_j$ too much on average.  For the notation in the
following corollary we refer to Section~\ref{sec:probdistr}, equation
(\ref{eq:29}) and the subsequent remarks.
\begin{corollary}\label{cor:disjointreps}
  Let~$\Pd_{T U^k V} := \Pd_{T}
  \Pd_{U_1|T}\Pd_{U_2|T}\dots\Pd_{U_k|T} \Pd_{V|T U^k}$
  be a probability distribution over~$\cT\times\cU^k \times \cV$,~$\Wevent$
  be an event. Then,
  \begin{align*}
    \sum_{j=1}^k 
    \Bigl\| \Pd_{T U_j V|\Wevent} - 
    \Pd_{T V|\Wevent}\Pd_{U_j|T} \Bigr\|
    &\leq \sqrt{k} \sqrt{\log(|\cV^*|) +
      \log\Bigl(\frac{1}{\Pr[\Wevent]}\Bigr)},
  \end{align*}
  where $\cV^* := \{ v \in \cV | \Pd_{V|W}(v) > 0\}$.
\end{corollary}
The proof is essentially an application of Jensen's inequality on
Lemma~\ref{lem:basicrepetitions}.
\begin{proof}
  Fix a pair $(t,v)\in\cT\times\cV$ and consider the
  distributions~$\Pd_{U^k|T=t,V=v,\Wevent}$ and $\Pd_{U^k|T=t}$.  We
  apply Lemma~\ref{lem:basicrepetitions} (in the form given by
  \eqref{eq:52}) on these distributions (with the
  event~$(V\mathord{=}v) \land \Wevent$) and get
  \begin{align}
    \sum_{j=1}^k 
    \Bigl\| \Pd_{T U_j V|\Wevent} \!-\! 
    \Pd_{T V|\Wevent}\Pd_{U_j|T} \Bigr\|
    &=\sum_{t \in \cT,v\in\cV^*} \Pd_{TV|\Wevent}(t,v) \cdot
    \sum_{j=1}^k \Bigl\| \Pd_{U_j|T=t, V=v,\Wevent} - \Pd_{U_j|T=t} \Bigr\|\nonumber\\
    &\leq
    \sum_{t \in \cT,v\in\cV^*} \Pd_{TV|\Wevent}(t,v)
    \sqrt{k \log\Bigl(\frac{1}{\Pr[\Wevent\land V=v|T=t]}\Bigr)}\nonumber\\
    &\leq
    \sqrt{k\log\Bigl(\sum_{t \in \cT,v\in\cV^*} \Pd_{TV|\Wevent}(t,v)
      \frac{1}{\Pr[\Wevent\land V=v|T=t]}\Bigr)},\label{eq:17}
  \end{align}
  where the last inequality is Jensen's inequality applied
  on the function~$\sqrt{\log(\cdot)}$ which is concave on~$[1,\infty)$.  We compute
  \begin{align}
    \sum_{t \in \cT,v\in\cV^*}\!\!\!\!\!\!
    \Pd_{TV|\Wevent}(t,v) \frac{1}{\Pr[W\land V=v|T=t]} 
    &=
    \sum_{t \in \cT,v\in\cV^*} \frac{\Pr[{T=t} \land
      {V=v}|\Wevent]}{\Pr[\Wevent\land V=v|T=t]} \nonumber\\
    &=
    \sum_{t \in \cT,v\in\cV^*} 
    \frac{\Pr[{T=t} \land {V=v} \land \Wevent] \Pr[T=t]}{\Pr[\Wevent]
      \Pr[V=v \land T=t \land \Wevent]} \nonumber\\
    &=
    \sum_{t \in \cT,v\in\cV^*}
    \frac{\Pr[T=t]}{\Pr[\Wevent]}
    =
    \frac{|\cV^*|}{\Pr[\Wevent]}.\nonumber
  \end{align}
  Inserting this into (\ref{eq:17}) completes the proof.
\end{proof}

\section{Embedding by Local Computation}
We next study under what conditions random variables can be embedded
into other random variables by local computations.
\begin{definition}[Embeddable]
  For two distributions~$\Pd_{X_0 Y_0}$ and $\Pd_{X_1 S
    Y_1 T}$ we say that $(X_0, Y_0)$ is $1-\eps$-embeddable in
  $(X_1 S, Y_1 T)$ with~$(X_1,Y_1)=(X_0,Y_0)$ if 
  there exists a probability measure~$\Pd_{R}$ over a set~$\cR$ and
  functions~$f_A: \cX\times\cR\rightarrow\cS$, $f_B:\cY\times\cR
  \rightarrow \cT$, such that
  \begin{align*}
    \bigl\| \Pd_{X_0Y_0}\Pd_{F_A F_B|XY}
    - \Pd_{X_1Y_1S T}\bigr\| \leq \eps,
  \end{align*}
  where~$\Pd_{F_A F_B|X=xY=y}$ is the distribution defined by the
  random variable~$(f_A(x,R),\n f_B(y,R))$.
\end{definition}
The following lemma gives a condition under which $(X,Y)$ is
embeddable in~$(XS,\n YS)$.  It is one of the main contributions of
this paper.
\begin{lemma}\label{lem:LocallyComputableCommonPartPre}
  Let a distribution~$\Pd_{SXY}$ be given.  If 
  \begin{align}
    \| \Pd_{SXY}-\Pd_{X Y}\Pd_{S|X}\| \leq \eps_1 \label{eq:12}
    \intertext{and}
    \| \Pd_{SXY}-\Pd_{X Y}\Pd_{S|Y}\| \leq \eps_2 ,\label{eq:13}
  \end{align}
  then~$(X, Y)$ is $1-2\eps_1-2\eps_2$-embeddable\footnote{It is
    understood that the embedding satisfies $(X,Y)=(X,Y)$, i.e., that
    the original random variables will result if from the resulting
    $(XS,YS)$ the $S$-part is omitted.}
  in~$(XS, YS)$.
\end{lemma}

Even if~$\eps_1=\eps_2=0$, equations (\ref{eq:12}) and (\ref{eq:13})
do not imply that~$S$ is independent of~$X$ and~$Y$. For example, if
$X$ and~$Y$ contain the
same uniform random bit, then $S$ can depend on this bit. 
However, if~$\eps_1=\eps_2 = 0$ the lemma is obviously true: Alice
uses shared randomness to choose~$S$ according to~$\Pd_{S|X=x}$ (more
concretely: Alice chooses a uniform random real~$\rho \in [0,1]$ and
uses the smallest element~$s$ for which the cumulative distribution
function~$\sum_{s'\leq s}\Pd_{S|X=x}(s')$ is larger than~$\rho$).
Since Bob has the same distribution~$\Pd_{S|Y=y}$ he will find the
same value if he uses the same shared randomness.

In case~$\eps_1>0$ and~$\eps_2>0$, we have to overcome the following
problem: $\Pd_{S|Y=y}$ is unknown to Alice (since~$y$ is unknown to
Alice), and analogously $\Pd_{S|X=x}$ is unknown to Bob.  The solution
is to define the function $f_A: \cX \times \cR \rightarrow \cS$ with
the following process: Alice chooses, using shared randomness, a
uniform random element~$s$ from $\cS$ and a uniform random real
number~$\rho\in[0,1]$.  If $\Pd_{S|X=x}(s)>\rho$ she outputs~$s$,
otherwise Alice repeats the above.  The function~$f_B: \cY \times \cR
\rightarrow \cS$ is defined by the analogous process given~$y$.  It is
easy to see that Alice outputs elements according to the
distribution~$\Pd_{S|X=x}$, Bob according to~$\Pd_{S|Y=y}$.  We
further show that usually the output of~$f_A$ is equal to the output
of~$f_B$.
\begin{proof}
  Let~$\cR := (\cS\times[0,1])^\infty$ be the set of infinite
  sequences over~$\cS\times[0,1]$.  For a fixed~$x,y$ and a sequence $r
  := \{(s_i,\rho_i)\}_{i\geq 0}$, we define~$f_A(x,r) := s_i$ if~$i$
  is the smallest index for which $\Pd_{S|X=x}(s_i) > \rho_i$.
  Analogously, $f_B(y,r) := s_j$ if~$j$ is the smallest index
  with~$\Pd_{S|Y=y}(s_j)>\rho_j$ and\footnote{The use of~$f_{AB}$ in
    order to simplify the analysis was suggested by Anup Rao.}
  $f_{AB}(x,y,r) := s_k$ if~$k$ is the smallest index
  with~$\Pd_{S|X=xY=y}(s_k)>\rho_k$.  If no such index exist the
  respective function is defined in an arbitrary way (this happens
  with probability~$0$).
  
  Let $\Pd_{XYF_AF_BF_{AB}}$ be the joint distribution
  of~$(x,y,f_A(x,r), f_B(y,r),f_{AB}(x,y,r))$ where $(x,y)$ is chosen
  according to $\Pd_{XY}$ and $r$ uniformly from $\cR$.  We have
  $\Pd_{F_{AB}|X=xY=y}=\Pd_{S|X=xY=y}$, $\Pd_{F_A|X=x}=\Pd_{S|X=x}$
  and $\Pd_{F_B|Y=y}=\Pd_{S|Y=y}$, since these equalities hold
  conditioned on the event that the respective function accepts in 
  round~$i$, for any fixed~$i$.
  
  Further, we have $\Pr[F_A=F_{AB}|X=x,Y=y] \geq
  1-2\|\Pd_{F_{A}|X=x}-\Pd_{F_{AB}|X=xY=y}\|$: the two values
  $F_A,F_{AB}$ are equal if $\rho_j <
  \min(\Pd_{F_{A}|X=x}(s_j),\Pd_{F_{AB}|X=xY=y}(s_j))$ for the
  smallest $j$ for which $\rho_j <
  \max(\Pd_{F_{A}|X=x}(s_j),\Pd_{F_{AB}|X=xY=y}(s_j))$ is satisfied. 
  This happens with probability
  \begin{align*}
    \frac{\sum_{s}\min(\Pd_{F_{A}|X=x}(s_j),\Pd_{F_{AB}|X=xY=y}(s_j))}
    {\sum_{s}\max(\Pd_{F_{A}|X=x}(s_j),\Pd_{F_{AB}|X=xY=y}(s_j))}
    &=\frac{1-\|\Pd_{F_{A}|X=x}-\Pd_{F_{AB}|X=xY=y}\|}{1+\|\Pd_{F_{A}|X=x}-\Pd_{F_{AB}|X=xY=y}\|}\\
    &\geq 1-2\|\Pd_{F_{A}|X=x}-\Pd_{F_{AB}|X=xY=y}\|.
  \end{align*}
  This yields $\Pr[F_A=F_{AB}] \geq 1-2\eps_1$, and analogously we get
  $\Pr[F_B=F_{AB}] \geq 1-2\eps_2$, and thus $\Pr[F_A=F_B=F_{AB}]\geq
  1-2\eps_1-2\eps_2$.  
  This implies
  \begin{align*}
    \|\Pd_{XYSS} - \Pd_{XY}\Pd_{F_AF_B|XY}\|
    &=
    \|\Pd_{XYF_{AB}F_{AB}} - \Pd_{XYF_AF_B}\| \geq 1-2\eps_1-2\eps_2.\qedhere
  \end{align*}
\end{proof}

In the following corollary, the input distribution is changed
slightly.  This makes it a bit easier to apply later.
\begin{corollary}\label{cor:LocallyComputableCommonPart}
  Let distributions~$\Pd_{SXY}$ and~$\Pd_{X_0 Y_0}$ be given.  If 
  \begin{align}
    \| \Pd_{SXY}-\Pd_{X_0 Y_0}\Pd_{S|X}\| \leq \eps_1 \label{eq:48}
    \intertext{and}
    \| \Pd_{SXY}-\Pd_{X_0 Y_0}\Pd_{S|Y}\| \leq \eps_2 ,\label{eq:49}
  \end{align}
  then~$(X_0, Y_0)$ is $1-3\eps_1-2\eps_2$-embeddable\footnote{The
    statement could be made symmetric (i.e., $(X_0,Y_0)$ is
    $1-2\eps_1-2\eps_2 - \min(\eps_1,\eps_2)$-embeddable).}
  in~$(XS, YS)$ with~$(X,Y)=(X_0,Y_0)$.
\end{corollary}
\begin{proof}
  From (\ref{eq:48}) we get $\| \Pd_{XY}-\Pd_{X_0Y_0}\|\leq\eps_1$.
  One can now find a joint distribution $\Pd_{XYX_0Y_0}$ with
  $\Pr[(X,Y)=(X_0,Y_0)] \geq 1-\eps_1$.  The corollary now follows by
  applying $f_A$ and~$f_B$ from
  Lemma~\ref{lem:LocallyComputableCommonPartPre}.
\end{proof}

Random variables~$S,T,U$ form a Markov chain,
written~$S\leftrightarrow T\leftrightarrow U$ if $\Pd_{STU} =
\Pd_{T}\Pd_{S|T}\Pd_{U|T}$ (i.e., if given~$T$ the probability
distribution of $U$ does not depend on~$S$).  The following lemma is
essentially Lemma~4.1 in Raz's paper.

\begin{lemma}\label{lem:locallyComputableMarkov}
  Let~$\Pd_{XYST}$ be any distribution. If 
  \begin{align*}
    S \leftrightarrow X \leftrightarrow YT 
    \intertext{and}
    XS \leftrightarrow Y \leftrightarrow T 
  \end{align*}
  then~$(X,Y)$ is $1$-embeddable in~$(XS,YT)$.
\end{lemma}
\begin{proof}
  Using individual (non-shared) randomness, Alice computes~$S$
  according to~$\Pd_{S|X=x}$ and Bob computes~$T$ according
  to~$\Pd_{T|Y=y}$.  Since
  \begin{align}
    \Pd_{STXY} = \Pd_{XY}\Pd_{S|XY}\Pd_{T|SXY} = \Pd_{XY}\Pd_{S|X}\Pd_{T|Y}
  \end{align}
  this gives the correct (global) distribution.
\end{proof}

\section{Embeddings for Games}
Given a game~$\mkG$ and its~$n$-fold parallel repetition, we now show
that $(X,Y)$ can be embedded into~$(\Xtilde^n,\Ytilde^n)$,
where~$\Pd_{\Xtilde^n\Ytilde^n} := \Pd_{X^nY^n|W_{k+1}\land\dots\land W_n}$.

We need the following simple fact on statistical distance.
\begin{fact}\label{fact:statdistsplit}
  Let~$\Pd_{Z_0}$ and~$\Pd_{Z_1}$ be distributions over~$\cZ$.
  Let $\cS \subseteq \cZ$ be such that $\Pr[Z_0 \in \cS] = \Pr[Z_1 \in
  \cS] = \frac{1}{2}$.
  Then, 
  \begin{align*}
    \| \Pd_{Z_0|Z_0 \in \cS} - \Pd_{Z_1|Z_1\in\cS}\|
    \leq 
    2\| \Pd_{Z_0} - \Pd_{Z_1}\|\, .
  \end{align*}
\end{fact}

Also, we need the following statements about Markov chains.

\begin{claim}\label{claim:hmm}
  Let~$\Pd_{X_0 Y_0}\Pd_{X_1Y_1}$ be a distribution
  over~$\cX_0\times\cY_0\times\cX_1\times\cY_1$, $f:
  \cX_0\times\cX_1\rightarrow \cU$ and~$g:
  \cY_0\times\cY_1\rightarrow\cV$ be arbitrary.  Then,
  \begin{align}\label{eq:22}
    X_0 X_1 \leftrightarrow X_0 f(X_0,X_1) Y_1 g(Y_0,Y_1)
    \leftrightarrow Y_0 Y_1.
  \end{align}
\end{claim}
\begin{proof}
  It is sufficient to show this for all possible values~$x_0\in\cX_0$
  and~$y_1\in\cY_1$.  Let~$\Pd_{\Ytilde_0\Xtilde_1} :=
  \Pd_{Y_0X_1|X_0=x_0 Y_1=y_1} = \Pd_{Y_0|X_0=x_0}\Pd_{X_1|Y_1=y_1}$.
  In this case, (\ref{eq:22}) reduces to
  \begin{align*}
    \Xtilde_1 \leftrightarrow
    f(x_0,\Xtilde_1) g(\Ytilde_0,y_1)\leftrightarrow \Ytilde_0.
  \end{align*}
  Since $\Xtilde_1$ and~$\Ytilde_0$ are independent this is obvious.
\end{proof}

\begin{claim}\label{claim:markovcondition}
  Let~$\Pd_{TUV}$ be a distribution
  over~$\cT\times\cU\times\cV$ and $W$ an event with
  \begin{align*}
    T &\leftrightarrow U \leftrightarrow V,\\
    W &\leftrightarrow U \leftrightarrow TV.
  \end{align*}
  Then, for~$\Pd_{\Ttilde\Utilde\Vtilde} := \Pd_{TUV|W}$ we have
  \begin{align*}
    \Ttilde\leftrightarrow \Utilde \leftrightarrow \Vtilde.
  \end{align*}
\end{claim}
\begin{proof}
  \begin{align*}
    \Pd_{\Ttilde\Utilde\Vtilde}(t,u,v) &= \Pd_{TUV|W}(t,u,v)\\
    &=\Pd_{U|W}(u)\Pd_{TV|U=u,W}(t,v)\\
    &=\Pd_{U|W}(u)\Pd_{TV|U=u}(t,v)\\
    &=\Pd_{U|W}(u)\Pd_{T|U=u}(t)\Pd_{V|U=u}(v)\\
    &=\Pd_{U|W}(u)\Pd_{T|U=u,W}(t)\Pd_{V|U=u,W}(v).\qedhere
  \end{align*}
\end{proof}

\begin{lemma}\label{lem:locallyComputableGame}
  Let a game $\mkG^{ n}=(Q^n, (\Pd_{XY})^n)$, a
  strategy $(h_a,h_b)$, and $k \leq n$ be given.  Let
  \begin{align*}
    \Pd_{\Xtilde^n\Ytilde^n} :=
    \Pd_{X^nY^n|W_{k+1}\land\dots\land W_n}
  \end{align*}
  
  Then, for~$1 \leq j \leq k$, there exists~$\eps_j \geq 0$ such
  that~$(X,Y)$ is~$1-\eps_j$-embeddable in~$(\Xtilde^n,\Ytilde^n)$
  with~$(\Xtilde_j,\Ytilde_j)=(X,Y)$ and
  \begin{align}
    \sum_{j=1}^{k}
    \eps_j \leq 
    15\sqrt{k}\sqrt{(n-k)\log(|\cA|\,|\cB|)+
      \log\Bigl(\frac{1}{\Pr[W_{k+1}\land\dots\land W_n]}\Bigr)}.\label{eq:27}
  \end{align}
\end{lemma}
\begin{proof}
  As described in Definition~\ref{def:randomvars} we consider the
  distribution~$\Pd_{X^nY^nA^nB^nW^n}$ and the corresponding random
  variables.  Additionally, we let~$D_{1}, \ldots, D_{k}$ be uniform
  and independent bits.  For~$1\leq j\leq k$ we define
  \begin{align*}
    U_j &:= \begin{cases}
      X_j & \text{if $D_j = 0$}\\
      Y_j & \text{otherwise}
    \end{cases}\\
    \intertext{and}
    \Ubar_j &:= \begin{cases}
      Y_j & \text{if $D_j = 0$}\\
      X_j & \text{otherwise.}
    \end{cases}
  \end{align*}
  Also, we set
  \begin{align}
    T &:= (X_{k+1},\ldots,X_{n}, Y_{k+1},\ldots,Y_n, D^k, \Ubar^k),\label{eq:44}\\
    V &:=      (A_{k+1},\ldots,A_{n},B_{k+1},\ldots,B_n),\label{eq:45}
  \end{align}
  and define the event~$W := W_{k+1}\land\dots \land W_n$.
  
  From Corollary~\ref{cor:disjointreps}  we get
  \begin{align}
    \sum_{j=1}^k 
    \Bigl\| \Pd_{T U_jV|W} - 
    \Pd_{T V|W}\Pd_{U_j|T} \Bigr\|
    &\leq \eps_{\Tot}\label{eq:14}\,,
  \end{align}
  where we set \begin{align} \eps_{\Tot} := \sqrt{k}
    \sqrt{(n-k)\log(|\cA||\cB|)+
      \log\Bigl(\frac{1}{\Pr[W]}\Bigr)}
  \end{align}
  (we applied Corollary~\ref{cor:disjointreps} using~$|\cV^*| \leq |\cV|$).

  In (\ref{eq:14}), we condition on both sides on the event~$D_j=0$,
  which is, on both sides, a restriction on a subset which has
  probability~$\frac{1}{2}$.  Fact~\ref{fact:statdistsplit} implies
  \begin{align}
    \sum_{j=1}^k 
    \Bigl\| \Pd_{T U_jV|W \land (D_j=0)} - 
    \Pd_{T V|W\land (D_j=0)}\Pd_{U_j|T} \Bigr\|
    &\leq 2\eps_{\Tot}\,,\label{eq:19}
  \end{align}
  where we do not need to condition on~$D_j=0$ in~$\Pd_{U_j|T}$ since
  this is included in the given~$t$ anyhow; in fact we can now
  write~$\Pd_{X_j|Y_j}$ instead of~$\Pd_{U_j|T}$.

  For a fixed~$j$, define the random variable
  \begin{align}
    T^\NoJ &:= 
    (X_{k+1},\ldots,X_{n}, Y_{k+1},\ldots,Y_n,\nonumber\\
    &\qquad\qquad D_1,\ldots,D_{j-1},D_{j+1},\ldots,D_k, \nonumber\\
    &\qquad\qquad\Ubar_1,\ldots,\Ubar_{j-1},\Ubar_{j+1},\ldots,\Ubar_{k}).
  \end{align}
  With this notation (\ref{eq:19}) is equivalent to
  \begin{align}
    \sum_{j=1}^k 
    \Bigl\| \Pd_{T^{\NoJ}X_j Y_j V|W \land (D_j=0)} - 
    \Pd_{T^{\NoJ} Y_j V|W \land (D_j=0)}\Pd_{X_j|Y_j} \Bigr\|
    &\leq 2\eps_{\Tot}\,.
  \end{align}
  But now nothing depends on~$D_j = 0$ anymore, so this also means
  \begin{align}\label{eq:4}
    \sum_{j=1}^k 
    \Bigl\| \Pd_{T^{\NoJ} X_j Y_j V|W} - 
    \Pd_{T^{\NoJ} Y_j V|W}\Pd_{X_j|Y_j} \Bigr\|
    &\leq 2\eps_{\Tot}\,.
  \end{align}
  We set~$S := (T^\NoJ,V)$ and define the probability distribution
  \begin{align}
    \Pd_{\Stilde \Xtilde^n \Ytilde^n} := \Pd_{S X^nY^n|W}.
  \end{align}
  With this, (\ref{eq:4}) becomes
  \begin{align}
    \sum_{j=1}^k 
    \Bigl\| \Pd_{\Stilde \Xtilde_j \Ytilde_j} - 
    \Pd_{\Stilde \Ytilde_j}\Pd_{X_j|Y_j} \Bigr\|
    &\leq 2\eps_{\Tot}\,,
  \end{align}
  or, equivalently
  \begin{align}
    \sum_{j=1}^k 
    \Bigl\| \Pd_{\Stilde \Xtilde_j \Ytilde_j} - 
    \Pd_{\Ytilde_j}\Pd_{\Stilde|\Ytilde_j}\Pd_{X|Y} \Bigr\|
    &\leq 2\eps_{\Tot}\,.
  \end{align}
  Lemma~\ref{lem:basicrepetitions} implies 
  \begin{align}
    \sum_{j=1}^k 
    \Bigl\| \Pd_{\Ytilde_j} - \Pd_{Y} \Bigl\| \leq \eps_{\Tot},
  \end{align}
  and thus
  \begin{align}\label{eq:25}
    \sum_{j=1}^k 
    \Bigl\| \Pd_{\Stilde \Xtilde_j \Ytilde_j} - 
    \Pd_{XY}\Pd_{\Stilde|\Ytilde_j} \Bigr\|
    &\leq 3\eps_{\Tot}\,.
  \end{align}
  Symmetric reasoning yields
  \begin{align}\label{eq:26}
    \sum_{j=1}^k 
    \Bigl\| \Pd_{\Stilde \Xtilde_j \Ytilde_j} - 
    \Pd_{XY}\Pd_{\Stilde|\Xtilde_j} \Bigr\|
    &\leq 3\eps_{\Tot}\,.
  \end{align}
  From~(\ref{eq:25}) and (\ref{eq:26}),
  Corollary~\ref{cor:LocallyComputableCommonPart} implies that~$(X,Y)$
  is $1-\eps_j$-embeddable in $(\Xtilde_j\Stilde,\Ytilde_j\Stilde)$
  with $(\Xtilde_j,\Ytilde_j)=(X,Y)$ and such that $\sum_{j=1}^k
  {\eps_j} \leq 15\eps_{\Tot}$.

  We next show that
  \begin{align}\label{eq:28}
    X^k \leftrightarrow T V \leftrightarrow Y^k.
  \end{align}
  If the bits~$D^k$ and the values $X_{k+1},\ldots,X_n$,
  $Y_{k+1},\ldots,Y_n$ are fixed, this follows immediately from
  Claim~\ref{claim:hmm}.  Since it holds for all these values it must
  also hold overall.

  From (\ref{eq:28}) we easily get
  \begin{align*}
    X^n \leftrightarrow X_j S \leftrightarrow
    Y^n Y_j S\\
    X^n X_j S \leftrightarrow Y_j S \leftrightarrow
    Y^n.
  \end{align*}
  Claim~\ref{claim:markovcondition} yields
  \begin{align}
    \Xtilde^n \leftrightarrow \Xtilde_j \Stilde \leftrightarrow
    \Ytilde^n\Ytilde_j\Stilde \label{eq:20}\\
    \Xtilde^n\Xtilde_j\Stilde \leftrightarrow \Ytilde_j\Stilde \leftrightarrow
    \Ytilde^n.\label{eq:21}
  \end{align}
  
  Above we have seen that $(X,Y)$ is embeddable
  in~$(\Xtilde_j\Stilde,\Ytilde_j\Stilde)$
  with~$(\Xtilde_j,\Ytilde_j)=(X,Y)$.  
  Lemma~\ref{lem:locallyComputableMarkov} together with (\ref{eq:20})
  and (\ref{eq:21}) now implies that we can $1$-locally embed this in
  $(\Xtilde^n\Xtilde_j\Stilde, \Ytilde^n\Ytilde_j\Stilde)$.  Since
  Alice and Bob can then ignore part of the constructed information
  this completes the proof.
\end{proof}

\begin{lemma}\label{lem:gameconditionedvalue}
  Let a game~$\mkG=(Q,\Pd_{XY})$, its~$n$-fold
  repetition~$\mkG^{n}$, and a strategy $(h_a,h_b)$
  for~$\mkG^{n}$ be given.  Let indices~$i_1,\ldots,i_m$ be
  given.  Then, there exists an index~$i_{m+1}$ such
  that
  \begin{align}\label{eq:3}
    \Pr[W_{i_{m+1}} | &W_{i_1}\land\dots\land W_{i_m}] \nonumber \\
    &\leq v(\mkG) +
    15\sqrt{\frac{1}{n-m}}\sqrt{m\log(|\cA||\cB|) +
      \log\Bigl(\frac{1}{\Pr[W_{i_1}\land\dots\land W_{i_m}]}\Bigr)}.
  \end{align}
\end{lemma}
\begin{proof}
  First, we can assume that the given indices $i_\ell$, $1\leq\ell\leq
  m$, are pairwise different (otherwise we get a stronger statement).
  Given this we can even assume that~$i_\ell = n-\ell+1$ by
  appropriately redefining the functions~$(h_a,h_b)$.
  
  Define the distribution~$\Pd_{\Xtilde^n \Ytilde^n} := \Pd_{X^n Y^n|
    W_{n-m+1}\land\cdots\land W_{n}}$.
  Lemma~\ref{lem:locallyComputableGame} implies that there exists an
  index $j$ such that~$(X,Y)$ is~$1-\eps$-embeddable
  in~$(\Xtilde^n,\Ytilde^n)$ with $(\Xtilde_j,\Ytilde_j)=(X,Y)$ and
  \begin{align*}
    \eps := 15\sqrt{\frac{1}{n-m}}\sqrt{m\log(|\cA||\cB|)+
      \log\Bigl(\frac{1}{\Pr[W_{n-m+1}\land\dots\land W_{n}]}\Bigr)}.
  \end{align*}
  Consider the following strategy for~$\mkG$.  On input~$(X,Y)$ Alice
  and Bob~$1-\eps$-embed this into~$(\Xtilde^n,\Ytilde^n)$
  with~$(\Xtilde_j,\Ytilde_j)=(X,Y)$.  Since the resulting
  distribution has statistical distance at most~$\eps$
  from~$\Pd_{\Xtilde^n \Ytilde^n}$, if they output coordinate~$j$
  of~$h_a(\Xtilde^n)$ and~$h_b(\Ytilde^n)$ they have probability at
  least~$\Pr[W_j | W_{n-m+1}\land\dots\land W_{n}]-\eps$ to win the
  initial game.  The shared randomness can be eliminated (see the
  remark after Definition~\ref{def:game}), and thus
  \begin{align*}
    v(\mkG) &\geq \Pr[W_j | W_{n-m+1}\land\dots\land
    W_{n}]-\eps.\qedhere
  \end{align*}
\end{proof}

\section{Parallel Repetition Theorem}
\begin{proof}[Proof (of Theorem~\ref{thm:pr_local})]
  Fix a strategy~$(h_a,h_b)$ for~$\mkG^n$.  Then, repeatedly choose
  the index~$i_{m+1}$ for which~$\Pr[W_{i_{m+1}}|W_{i_1}\land \dots
  \land W_{i_{m}}]$ is minimized.  We set $p_0 := 1$ and $p_m :=
  \Pr[W_{i_1}\land \dots \land W_{i_{m}}]$.
  Lemma~\ref{lem:gameconditionedvalue} implies
  \begin{align}
    p_{m+1} \leq  p_{m} \cdot \biggl(v+15\sqrt{\frac{1}{n-m}}\sqrt{m\log(|\cA||\cB|) +
      \log\Bigl(\frac{1}{p_{m}}\Bigr)}\biggr).\label{eq:33}
  \end{align}
  We show per induction that
  \begin{align*}
    p_m \leq \Bigl(\frac{1+v}{2}\Bigr)^{m},
  \end{align*}
  as long as $m \leq \frac{(1-v)^2 (n-m)}{2700\log(|\cA||\cB|)}$.  The
  statement holds for~$m=0$ and we now make a step from~$m$ to~$m+1$.
  First, we can assume that~$p_{m} \geq
  \bigl(\frac{1+v}{2}\bigr)^{m+1} > \frac{1}{2}^{m+1}$, as otherwise
  the induction step is trivial.  In this case, \eqref{eq:33} yields
  \begin{align}
    p_{m+1} 
    &\leq  
    p_{m} \cdot \biggl(v+15\sqrt{\frac{1}{n-m}}\sqrt{m\log(|\cA||\cB|) +
      (m+1)}\biggr)\nonumber\\
    & \leq
    p_{m} \cdot \biggl(v+\sqrt{\frac{1}{n-m}}\sqrt{675 m\log(|\cA||\cB|)}\biggr)\label{eq:35}    
  \end{align}
  Since we assume~$m \leq \frac{(1-v)^2}{2700
    \log(|\cA||\cB|)}(n-m)$ this proves the induction step.

  In total we get for~$m = \frac{n(1-v)^2}{3000\log(|\cA||\cB|)}$
  \begin{align}\label{eq:38}
    p_{m} \leq \Bigl(\frac{1+v}{2}\Bigr)^\frac{n(1-v)^2}{3000\log(|\cA||\cB|)}.
  \end{align}
  We have
  \begin{align}
    \Bigl(\frac{1+v}{2}\Bigr)^{\frac{(1-v)^2}{3000}}
    &= \Bigl(1-\frac{1-v}{2}\Bigr)^{\frac{(1-v)^2}{3000}} \nonumber\\
    &\leq 1-\frac{(1-v)^3}{6000},\label{eq:39}
  \end{align}
  where the last inequality follows from 
  $(1-b)^a \leq 1-ab$ which holds for all $a \in [0,1]$, $b\leq1$. 
  Since~$\Pr[W_1\land\dots\land W_n] \leq p_m$, (\ref{eq:38}) and
  (\ref{eq:39}) imply the theorem.\footnote{The minimal value of the
    sequence defined by $p_0 := 1$ and $p_{m+1} := p_m
    \bigl(v+\sqrt{\frac{225}{n-m}}\sqrt{m \ell + \log(1/p_m)}\bigr)$
    is indeed $\Bigl(1-\Theta((1-v)^3)\Bigr)^{\frac{n}{\ell}}$.  The
    argument in the proof above shows that the minimal value can 
    only be lower. On the other hand, the sequence given by $p'_0 := 1$,
    $p'_{m+1} := p'_m\Bigl(v+\sqrt{\frac{m\ell}{n}}\Bigr)$
    is strictly smaller than the sequence $\{p_j\}_{j\geq0}$.  This
    sequence does not decrease anymore if~$m > m ' := n(1 - v)^2
    /\ell$, and
    \begin{align*}
  p'_{m'} &= \prod_{i=0}^{m'-1} \Bigl(v+\sqrt{\frac{i\ell}{n}}\Bigr) 
  =
  \exp\left(\sum_{i=0}^{m'-1} \ln\Bigl(v+\sqrt{\frac{i\ell}{n}}\Bigr)\right)\\
  &\approx
  \exp\left(\int_{0}^{m'} \ln\Bigl(v+\sqrt{\frac{i\ell}{n}}\Bigr)\right)
  =
  \exp\Bigl((4v-1+2v^2 \ln(v) -
  3v^2)\cdot\frac{n}{2\ell}\Bigr)
  \approx
  \Bigl(1-\frac{(1-v)^3}{2}\Bigr)^{\frac{3n}{4\ell}}.
\end{align*}}
\end{proof}

\section{Improving the Rate}\label{sec:reducingtheexponent}

Theorem~\ref{thm:pr_local} shows that the~$n$-fold parallel repetition
reduces the winning probability from~$v(\mkG)$ to
$(1-\Theta(1-v(\mkG))^3)^{\Omega(\frac{n}{\log(|\cA||\cB|)})}$.  As
shown in \cite{PaRaWi97}, the term~$|\cA|\cdot|\cB|$ in the exponent
can be reduced to the the maximum (over~$x$, $y$) number of
(fractional) rectangles needed to cover the $1$-entries
in~$Q(x,y,\cdot,\cdot)$.  Here, we show that it can be reduced to a
quantity which is possibly smaller in some cases.

\begin{definition}[Exact Fractional Product Cover]
  Let~$Q: \cA\times\cB\rightarrow\{0,1\}$ be an arbitrary predicate.
  Two functions $f: \cA \times \{1,\ldots,\alpha\} \rightarrow [0,1]$
  and~$g: \cB\times \{1,\ldots,\alpha\}\rightarrow [0,1]$ form an
  \emph{exact fractional product cover of size~$\alpha$} for~$Q$ if for
  all~$a,b$:
  \begin{align*}
    Q(a,b) = \sum_{i=1}^\alpha f(a,i)g(b,i).
  \end{align*}
\end{definition}

Clearly, any partition by rectangles  gives an exact fractional
product cover (by definining $f(a,i)$ and~$g(b,i)$ as appropriate
predicates).  We will prove the following strengthening of
Theorem~\ref{thm:pr_local}.
\begin{theorem}\label{thm:pr_local_strengthened}
  Let~$\mkG = (\Pd_{XY},Q)$ be a game.  Let~$\alpha$ be such that for
  all~$(x,y)$ there exists an exact fractional product cover of size~$\alpha$
  for~$Q_{x,y}(a,b) := Q(x,y,a,b)$.
  If~$\alpha > 1$ then
  \begin{align}
    v(\mkG^{ n}) \leq
    \Bigl(1-\frac{(1-v)^3}{6000}\Bigr)^{\frac{n}{\log(\alpha)}},\label{eq:23}
  \end{align}
  and if~$\alpha = 1$ then
  \begin{align}
    v(\mkG^{ n}) \leq
    \Bigl(1-\frac{(1-v)^2}{6000}\Bigr)^{n}.\label{eq:24}
  \end{align}
\end{theorem}

To prove Theorem~\ref{thm:pr_local_strengthened} we first need a
characterization of fractional product covers by Markov chains.

\begin{lemma}\label{lem:fractionalCovEq}
  Let a distribution $\Pd_{ABZ}=\Pd_{A}\Pd_{B}\Pd_{Z|AB}$ be given for which
  there exists functions $f(a,z): \cA\times\cZ \rightarrow [0,1]$
  and~$g(b,z): \cB\times\cZ\rightarrow[0,1]$ which satisfy
  \begin{align}\label{eq:9}
    \Pd_{Z|A=a B=b}(z) = f(a,z)\cdot g(b,z).
  \end{align}
  Then, $A\leftrightarrow Z \leftrightarrow B$.
\end{lemma}
Lemma~\ref{lem:fractionalCovEq} could be strengthened as follows: if
$\Pd_{Z|AB}$ is such that $A\leftrightarrow Z\leftrightarrow B$ for
all distributions $\Pd_{A}\Pd_{B}$, then $\Pd_{Z|AB}$ is of the form
(\ref{eq:9}) for some functions $f$ and~$g$.  For completeness, we
prove this in Appendix~\ref{app:fractEquivalence}.

Lemma~\ref{lem:fractionalCovEq} implies the following: if~$Q:
\cA\times\cB \rightarrow \{0,1\}$ has a fractional product cover of
size~$\alpha$, then there exists a random variable~$Z$ over some
set~$\cZ$ given by a conditional distribution~$\Pd_{Z|AB}$ with the
following properties:
\begin{itemize}
\item For any product distribution~$\Pd_{AB}=\Pd_{A}\Pd_{B}$ we
  have~$A \leftrightarrow Z \leftrightarrow B$
\item $|\{z \in \cZ|\exists a,b: Q(a,b)=1 \land \Pd_{Z|A=aB=b}(z) >
  0\}| \leq \alpha$
\item $Q(a,b)$ can be inferred from~$z$.
\end{itemize}
(Note that we do not restrict the alphabet size of~$Z$ in
case~$Q(a,b)=0$, which means that in this case $z$ can be, for example,
$(a,b)$.)

\begin{proof}[Proof (of Lemma~\ref{lem:fractionalCovEq})]
  We get
  \begin{align*}
    \Pd_{A|B=b Z=z}(a) 
    &=
    \frac{\Pd_{ABZ}(a,b,z)}{\Pd_{BZ}(b,z)}\\
    &=
    \frac{\Pd_{A}(a)\Pd_{B}(b)f(a,z)g(b,z)}
    {\sum_{a'}\Pd_{B}(b)\Pd_{A}(a')f(a',z)g(b,z)}\\
    &=
    \frac{\Pd_{A}(a)f(a,z)}
    {\sum_{a'}\Pd_{A}(a')f(a',z)}\\
    &=\frac{\sum_{b'} \Pd_{A}(a)\Pd_{B}(b')f(a,z)g(b',z)}
    {\sum_{a',b'}\Pd_{A}(a')\Pd_{B}(b')f(a',z)g(b',z)}\\
    &=\frac{\Pd_{AZ}(a,z)}{\Pd_{Z}(z)}\\
    &=\Pd_{A|Z=z}(a),
  \end{align*}
  and thus $\Pd_{ABZ}(a,b,z) = \Pd_{Z}(z)\Pd_{B|Z=z}(b)
  \Pd_{A|B=bZ=z}(a) = 
  \Pd_{Z}(z)\Pd_{B|Z=z}(b)
  \Pd_{A|Z=z}(a)$, which means that $A\leftrightarrow
  Z \leftrightarrow B$.
\end{proof}

Given the characterization from Lemma~\ref{lem:fractionalCovEq} we can
now prove Theorem~\ref{thm:pr_local_strengthened}.

\begin{proof}[Proof (of Theorem~\ref{thm:pr_local_strengthened})]
  We first show that Lemma~\ref{lem:locallyComputableGame} still holds
  if we replace (\ref{eq:27}) by
  \begin{align}
    \sum_{j=1}^{k}
    \eps_j \leq 
    15\sqrt{k}\sqrt{(n-k)\log(\alpha)+
      \log\Bigl(\frac{1}{\Pr[W_{k+1}\land\dots\land W_n]}\Bigr)}.\label{eq:34}
  \end{align}
  
  For this, we define the random variables~$D^k$, $U^k$, $\Ubar^k$,
  and $T$ exactly as in the proof of
  Lemma~\ref{lem:locallyComputableGame}.  Instead of (\ref{eq:45}) we
  now define
  \begin{align}
    V := (Z_{k+1},\ldots,Z_{n}),
  \end{align}
  where $Z_i$ is obtained from $(A_i,B_i,X_i,Y_i)$ by a channel that
  has alphabet size at most~$\alpha$ in case~$W_i$, which ensures $A_i
  \leftrightarrow X_iY_iZ_i\leftrightarrow B_i$ in case $A_i$
  and~$B_i$ are independent, and for which $W_i$ can be inferred from
  $(X_i,Y_i,Z_i)$.  The existence of such a random variable is ensured
  by Lemma~\ref{lem:fractionalCovEq} and the fact that for
  every~$(x,y)$ there exists a exact fractional product cover of
  size~$\alpha$ for~$Q(x,y,\cdot,\cdot)$ (the alphabet size of~$Z$ in
  case $Q(x,y,a,b)=0$ is irrelevant and $Z$ can be defined, for
  example, as~$(A,B)$ in this case).
  
  From Corollary~\ref{cor:disjointreps} we now get
  \begin{align}
    \sum_{j=1}^k 
    \Bigl\| \Pd_{T U_jV|W} - 
    \Pd_{T V|W}\Pd_{U_j|T} \Bigr\|
    &\leq \eps_{\Tot}\label{eq:14b}\,,
  \end{align}
  where we set
  \begin{align} \eps_{\Tot} := \sqrt{k}
    \sqrt{(n-k)\log(\alpha)+
      \log\Bigl(\frac{1}{\Pr[W]}\Bigr)}.
  \end{align}
  
  For a fixed~$j$ we define $T^\NoJ$ as in the proof of
  Lemma~\ref{lem:locallyComputableGame} and obtain in exactly the same
  way for $S := (T^\NoJ,V)$ and
  \begin{align}
    \Pd_{\Stilde \Xtilde^n \Ytilde^n} := \Pd_{S X^nY^n|W}
  \end{align}
  the equations
  \begin{align}
    \sum_{j=1}^k 
    \Bigl\| \Pd_{\Stilde \Xtilde_j \Ytilde_j} - 
    \Pd_{XY}\Pd_{\Stilde|\Ytilde_j} \Bigr\|
    &\leq 3\eps_{\Tot}
  \end{align}
  and
  \begin{align}
    \sum_{j=1}^k 
    \Bigl\| \Pd_{\Stilde \Xtilde_j \Ytilde_j} - 
    \Pd_{XY}\Pd_{\Stilde|\Xtilde_j} \Bigr\|
    &\leq 3\eps_{\Tot}\,.
  \end{align}
  Again, 
  Corollary~\ref{cor:LocallyComputableCommonPart} implies that~$(X,Y)$
  is $1-\eps_j$-embeddable in $(\Xtilde_j\Stilde,\Ytilde_j\Stilde)$
  with $(\Xtilde_j,\Ytilde_j)=(X,Y)$ and such that $\sum_{j=1}^k
  {\eps_j} \leq 15\eps_{\Tot}$.

  Again we get
  \begin{align}
    X^k \leftrightarrow T V \leftrightarrow Y^k,
  \end{align}
  now using the properties of the~$Z_i$. (This is done as follows:
  clearly, $X^k \leftrightarrow T \leftrightarrow Y^k$, i.e. for a
  fixed values~$t$ for~$T$ the $X^k$ and $Y^k$ are independent.  Now,
  inductively adding $Z_i$ will not change this in any step.)
  Claim~\ref{claim:markovcondition} now yields
  \begin{align}
    \Xtilde^n \leftrightarrow \Xtilde_j \Stilde \leftrightarrow
    \Ytilde^n\Ytilde_j\Stilde \label{eq:42}\\
    \Xtilde^n\Xtilde_j\Stilde \leftrightarrow \Ytilde_j\Stilde \leftrightarrow
    \Ytilde^n,\label{eq:43}
  \end{align}
  and Lemma~\ref{lem:locallyComputableMarkov} completes the proof that
  (\ref{eq:34}) can replace (\ref{eq:27}) in
  Lemma~\ref{lem:locallyComputableGame}.

  From Lemma~\ref{lem:locallyComputableGame} where (\ref{eq:27}) is
  replaced by~(\ref{eq:34}) we obtain~(\ref{eq:23}) exactly as in the
  proof of Theorem~\ref{thm:pr_local}.  To get~(\ref{eq:24}) we note
  first that in this case (\ref{eq:34}) reduces to
  \begin{align}\label{eq:46}
    \sum_{j=1}^{k}
    \eps_j \leq 
    15\sqrt{k}\sqrt{\log\Bigl(\frac{1}{\Pr[W_{k+1}\land\dots\land W_n]}\Bigr)}.
  \end{align}
  Using an analogous definition for~$p_m$ as previously, we get
  \begin{align}\label{eq:6}
    p_{m+1} \leq p_{m} \cdot
    \Bigl(v+15\sqrt{\frac{1}{n-m}\log\Bigl(\frac{1}{p_{m}}\Bigr)}\Bigr).
  \end{align}
  Here, we show per induction that $p_m \leq (\frac{1+v}2)^m$ as long as
  $m+1 \leq \frac{(n-m)(1-v)}{400}$.  To make a step from~$m$ to~$m+1$
  we can assume~$p_m \geq (\frac{1+v}2)^{m+1}$, which implies $p_m
  \geq 2^{-(1-v)(m+1)}$ (since $(1-\frac12)^{1-v} \leq
  1-\frac{1-v}{2}$, see inequality below), which means that
  \begin{align}
    p_{m+1} &\leq p_{m}
    \cdot\Bigl(v+\sqrt{\frac{225(m+1)(1-v)}{n-m}}\Bigr),
  \end{align}
  for relevant values of~$m$.  If $m+1 \leq \frac{(n-m)(1-v)}{900}$
  this implies the hypothesis.  We thus get for~$m =
  \frac{n(1-v)}{1800}$
  \begin{align*}
    p_m &\leq
    \Bigl(\frac{1+v}{2}\Bigr)^{\frac{n(1-v)}{1800}}.
  \end{align*}
  Finally, 
  $(\frac{1+v}{2})^{(1-v)/1800} = (1-\frac{(1-v)}{2})^{(1-v)/1800}
  \leq 1-\frac{(1-v)^2}{3600}$,
  again using $(1-b)^a\leq 1-ab$ for $a\in[0,1]$, $b\leq
  1$.\footnote{Note that in case $p_m \leq
    \Bigl(1-\Theta((1-v)^2)\Bigr)^{n}$ equation~(\ref{eq:6})
    only implies $p_{m+1} \leq 
    p_m (v + 15\sqrt{-\log(1-\Theta((1-v)^2))}) =
    p_m (v + 15\sqrt{\Theta((1-v)^2)}) = 
    p_m (v + 1-v) = p_m$,
    and thus (\ref{eq:6}) cannot be
    used to get a significantly stronger version of the theorem.}
\end{proof}

\section{No-signaling Strategies}\label{sec:nosignal}

No-signaling strategies are those where the only restriction on the
response of Alice and Bob is that they do not \emph{imply}
communication.

\begin{definition}[No-signaling]
  A pair~$(h_a,h_b)$ of functions is \emph{no-signaling} if~$h_a:
  \cX\times\cY\times\cR \rightarrow \cA$ and~$h_b:
  \cX\times\cY\times\cR\rightarrow\cB$ satisfy
  \begin{align*}
    \Pr_{R}[h_a(x,y,R)] &= \Pr_R[h_a(x,y',R)]\\
    \Pr_{R}[h_b(x,y,R)] &= \Pr_R[h_b(x',y,R)],
  \end{align*}
  for all $x,x',y,y'$.
\end{definition}
\begin{definition}[No-signaling value]
  The no-signaling value $\vns(\mkG)$ of a
  \emph{game~$\mkG=(\Pd_{XY},Q)$
    over~$\cX\times\cY\times\cA\times\cB$} is
  \begin{align*}
    \vns(\mkG) := \max \Pr_{XYR}[Q(X,Y,h_a(X,Y,R),h_b(X,Y,R))],
  \end{align*}
  where the maximum is over all no-signaling functions $(h_a,h_b)$.
\end{definition}
Clearly,~$v(\mkG) \leq \vns(\mkG)$, since any local strategy is a
no-signaling strategy.  We further note that for no-signaling
strategies $\vns(\mkG^2) > (\vns(\mkG))^2$ is also possible, similar to
the local case (see Appendix~\ref{app:nontriviality}).

We will prove the following Theorem:
\begin{theorem}\label{thm:pr_nosignaling}
  For any game~$\mkG$ with no-signaling value~$\vns := \vns(\mkG)$ and any
  integer~$n$:
  \begin{align}\label{eq:7}
    \vns(\mkG^n) \leq 
    \Bigl(1-\frac{(1-\vns)^2}{6400}\Bigr)^{n}. 
  \end{align}
\end{theorem}
We remark that the proof of this theorem will be much simpler than the
proof of Theorem~\ref{thm:pr_local}.

We first show that if $\Pd_{XYST}$ a distribution which is close to
no-signalling (i.e., $\|\Pd_{XYS} - \Pd_{XY}\Pd_{S|X}\|$ and
$\Pd_{XYT}-\Pd_{XY}\Pd_{S|Y}$) then there exists a no-signalling
strategy which produces value which are statistically close to~$S$
and~$T$ from~$X$ and~$Y$.

\begin{lemma}\label{lem:no-signaling-close-helper}
  Let~$\Pd_{ST}$, $\Pd_{S'}$ be arbitrary distributions
  over~$\cS\times\cT$ and~$\cS$,~$\cS$ and~$\cT$ finite.
  Then, there exists a distribution~$\Pd_{\Sbar\,\Tbar}$ such that
  \begin{align}
    \| \Pd_{\Sbar\,\Tbar} - \Pd_{ST} \| &\leq \| \Pd_{S'} - \Pd_{S} \|\label{eq:30}\\
    \| \Pd_{\Sbar} - \Pd_{S'}\| &= 0\label{eq:31}\\
    \| \Pd_{\Tbar} - \Pd_{T} \| &= 0.\label{eq:32}
  \end{align}
\end{lemma}
\begin{proof}
  We change~$\Pd_{ST}$ gradually to~$\Pd_{\Sbar\,\Tbar}$
  such that in the end~(\ref{eq:31}) and (\ref{eq:32}) hold.
  
  For this, fix values~$s_0$ and~$s_1$ with
  \begin{align}\label{eq:18}
    \Pd_{S}(s_0) < \Pd_{S'}(s_0) \text{ and }
    \Pd_{S}(s_1) > \Pd_{S'}(s_1).  
  \end{align}
  Then, as long as~(\ref{eq:18}) holds find a value~$t$ for
  which~$\Pd_{ST}(s_1,t) > 0$.  Decrease $\Pd_{ST}(s_1,t)$ by~$\eps$
  and increase $\Pd_{ST}(s_0,t)$ by $\eps$, such that afterwards
  $\Pd_{ST}(s_1,t) = 0$ or (\ref{eq:18}) does not hold anymore
  for~$s_0$, $s_1$.  After a finite number of repetitions (\ref{eq:18})
  is not true anymore, and we start the process over again with new
  values for~$s_0,s_1$.  However, this can also only happen a finite
  number of times, thus the process terminates.
  
  If~(\ref{eq:18}) cannot be satisfied then clearly~(\ref{eq:31})
  holds.  We never change~$\Pd_{T}(t)$ for any~$t$ which
  implies~(\ref{eq:32}).  Finally, (\ref{eq:30}) is ensured by the
  fact that we only decrease~$\|\Pd_{S'}-\Pd_{S}\|$ and do not change
  $\Pd_{ST}$ more than~$\Pd_{S}$.
\end{proof}

\begin{lemma}\label{lem:nosignalingclose}
  Let~$\Pd_{X_0Y_0}$ and $\Pd_{XYST}$ be arbitrary distributions.  If
  \begin{align}
    \| \Pd_{X_0Y_0}\Pd_{S|X} - \Pd_{XYS} \| \leq \eps_1,\\
    \| \Pd_{X_0Y_0}\Pd_{T|Y} - \Pd_{XYT} \| \leq \eps_2,
  \end{align}
  then there exists a conditional
  distribution~$\Pd_{S'T'|X'=xY'=y}$ with $\Pd_{S'|X'=x Y'=y} =
  \Pd_{S'|X'=x}$ and $\Pd_{T'|X'=xY'=y} = \Pd_{T'|Y'=y}$ such
  that
  \begin{align}
    \| \Pd_{X_0Y_0}\Pd_{S'T'|XY} - \Pd_{XYST} \| \leq 3\eps_1 + 2\eps_2.
  \end{align}
\end{lemma}
\begin{proof}
  For fixed~$x,y$ we define~$\Pd_{S_0T_0|X=xY=y}$ using
  Lemma~\ref{lem:no-signaling-close-helper} with the following properties:
  \begin{align*}
   \|\Pd_{S_0T_0|X=xY=y}-\Pd_{ST|X=xY=y}\|&\leq\|\Pd_{S|X=x}-\Pd_{S|X=xY=y}\|\\
   \|\Pd_{S_0|X=xY=y}-\Pd_{S|X=x}\| & = 0\\
   \|\Pd_{T_0|X=xY=y}-\Pd_{T|X=xY=y}\| &= 0.
  \end{align*}
  Then, again using Lemma~\ref{lem:no-signaling-close-helper} we
  define $\Pd_{S'T'|X=xY=y}$ such that
  \begin{align*}
    \|\Pd_{S'T'|X=xY=y}-\Pd_{S_0T_0|X=xY=y}\| &\leq 
    \|\Pd_{T_0|Y=y}-\Pd_{T_0|X=xY=y}\|\\
    \|\Pd_{T'|X=xY=y}-\Pd_{T_0|Y=y}\| &= 0\\
    \|\Pd_{S'|X=xY=y}-\Pd_{S_0|X=xY=y}\| & = 0.
  \end{align*}
  We see that for all pairs~$x,y$ we have $\Pd_{S'|X=xY=y} =
  \Pd_{S'|X=x}$ and $\Pd_{T'|X=xY=y}=\Pd_{T'|Y=y}$.
  
  We further get
  \begin{align*}
    \|\Pd_{X_0Y_0}& \Pd_{S'T'|XY}-\Pd_{XYST}\| \\
    &\leq
    \eps_1+\|\Pd_{XY}\Pd_{S'T'|XY} - \Pd_{XYST}\|\\
    &=
    \eps_1+\sum_{x,y,s,t} \bigl|\Pd_{XY}(x,y)\Pd_{S'T'|X=xY=y}(s,t) 
    - \Pd_{XY}(x,y)\Pd_{ST|X=xY=y}(s,t) \bigr|\\
    &\leq
    \eps_1+\sum_{x,y}\Pd_{XY}(x,y)\Bigl( \| \Pd_{S|X=x}-\Pd_{S|X=xY=y}\| +
    \|\Pd_{T|Y=y}-\Pd_{T|X=xY=y}\|\Bigr)\\
    &\leq
    \eps_1+\|\Pd_{XY}\Pd_{S|X}-\Pd_{SXY}\| +
    \|\Pd_{XY}\Pd_{T|Y}-\Pd_{TXY}\|\\
    &\leq
    3\eps_1+2\eps_2.\qedhere
  \end{align*}
\end{proof}

We can now prove a non-signaling analogue of
Lemma~\ref{lem:gameconditionedvalue}.\footnote{A previous version of
  the proof of this lemma contained an error, which was first noticed
  by Oded Regev and Ricky Rosen.}
\begin{lemma}\label{lem:ns_gameconditionedvalue}
  Let a game~$\mkG=(Q,\Pd_{XY})$, its~$n$-fold repetition~$\mkG^{n}$,
  and a no-signaling strategy $(h_a,h_b)$ for~$\mkG^{n}$ be given.  Let
  indices~$i_1,\ldots,i_m$ be given.  Then, there exists an
  index~$i_{m+1}$ such that
  \begin{align}
    \Pr[W_{i_{m+1}} | &W_{i_1}\land\dots\land W_{i_m}] \nonumber \\
    &\leq \vns(\mkG) +
    10\sqrt{\frac{1}{n-m}}\sqrt{\log\Bigl(\frac{1}{\Pr[W_{i_1}\land\dots\land W_{i_m}]}\Bigr)}.
  \end{align}
\end{lemma}
\begin{proof}
  As in the proof of Lemma~\ref{lem:gameconditionedvalue} we assume
  that $i_\ell = n-\ell+1$ and we define~$W :=
  W_{n-m+1}\land\dots\land W_{n}$.  The no-signaling property of
  $(h_a, h_b)$ implies $\Pd_{X^nY^nA^n} =
  \Pd_{X^n}\Pd_{Y^n|X^n}\Pd_{A^n | X^n} = \Pd_{A^nX^n}\Pd_{Y^n|X^n}$.
  Thus, when we apply Corollary~\ref{cor:disjointreps} on this
  distribution (with the event~$W$ and the random variables $T=(X^n,
  A^n)$ and $U_j = Y_j$) we get
  \begin{align*}
    \sum_{j=1}^{n-m}\Bigl\| \Pd_{X^nA^nY_j|W} -
    \Pd_{X^nA^n|W}\Pd_{Y_j|X_j}\Bigr\| &=
    \sum_{j=1}^{n-m}\Bigl\| \Pd_{T Y_j|W} - \Pd_{T|W}\Pd_{Y_j|T} \Bigr\|\\
    &\leq \sqrt{(n-m)\log\Bigl(\frac{1}{\Pr[W]}\Bigr)}.
  \end{align*}
  Taking appropriate marginals this gives
  \begin{align*}
   \sum_{j=1}^{n-m}\Bigl\| \Pd_{X_jY_jA_j|W} -
    \Pd_{X_jA_j|W}\Pd_{Y_j|X_j}\Bigr\| 
    \leq \sqrt{(n-m)\log\Bigl(\frac{1}{\Pr[W]}\Bigr)}.
   \end{align*}
  Applying Lemma~\ref{lem:basicrepetitions} once more and rearranging
  we get
  \begin{align}\label{eq:15}
   \sum_{j=1}^{n-m}\Bigl\| \Pd_{X_jY_jA_j|W} -
    \Pd_{XY}\Pd_{A_j|X_j W}\Bigr\| 
    \leq 2\sqrt{(n-m)\log\Bigl(\frac{1}{\Pr[W]}\Bigr)}.
  \end{align}
  Symmetrically, we obtain
  \begin{align}\label{eq:11}
    \sum_{j=1}^{n-m} \Bigl\| 
    \Pd_{X_j Y_j B_j|W} - 
    \Pd_{XY}\Pd_{B_j|Y_j W}
    \Bigr\| \leq 
    2\sqrt{(n-m)\log\Bigl(\frac{1}{\Pr[W]}\Bigr) }.
  \end{align}
  From (\ref{eq:15}), (\ref{eq:11}), and
  Lemma~\ref{lem:nosignalingclose} we get that there exists a
  distribution $\Pd_{A_j'B_j'|XY}$ which can be implemented by
  no-signaling functions and for which
  \begin{align*}
    \sum_{j=1}^{n-m}
    \Bigl\| 
    \Pd_{XY}\Pd_{A_j'B_j'|XY} -
    \Pd_{X_jY_jA_jB_j|W}
    \Bigr\| \leq 10
    \sqrt{(n-m)\Bigl(\log\Bigl(\frac{1}{\Pr[W]}\Bigr)\Bigr)}
      .
  \end{align*}
  Thus, if Alice and Bob use the strategy implied by~$\Pd_{A_j'B_j'|XY}$
  (which is no-signaling) they can win the initial game with
  probability $\Pr[W_j|W] - 10\sqrt{1/(n-m)}\sqrt{\log(1/\Pr[W])}$ for
  some~$j$, which implies the lemma.
\end{proof}

\begin{proof}[Proof (of Theorem~\ref{thm:pr_nosignaling})]
  Fix a no-signaling strategy~$(h_a,h_b)$ for~$\mkG$.  As in the proof
  of Theorem~\ref{thm:pr_local} we repeatedly select indices~$i_{m+1}$
  such that~$\Pr[W_{i_{m+1}}|W_{i_1}\land\dots\land W_{i_m}]$ is
  minimized.  Let~$p_m := \Pr[W_{i_1}\land\dots\land W_{i_m}]$.
  Lemma~\ref{lem:ns_gameconditionedvalue} implies
  \begin{align}\label{eq:36}
    p_{m+1} \leq p_{m} \cdot
    \Bigl(v+10\sqrt{\frac{1}{n-m}\log\Bigl(\frac{1}{p_{m}}\Bigr)}\Bigr).
  \end{align}
  From (\ref{eq:36}) we obtain (\ref{eq:7}) in the same way as we
  obtained (\ref{eq:24}) from (\ref{eq:6}) in the proof of
  Theorem~\ref{thm:pr_local_strengthened}.
\end{proof}

\section*{Acknowledgments}
I would like to thank Paul Beame, Georges Baatz, Ryan O'Donnell, Johan
H\aa{}stad, Bartosz Przydatek, Anup Rao, Ran Raz, Renato Renner, Ricky
Rosen, Stefano Tessaro, Stefan Wolf, and J\"urg Wullschleger for
helpful discussions.  Paul Beame pointed out \cite{PaRaWi97} to me and
explained how the strengthening given there can be adapted to the
proof given here; he also simplified an argument in a previous version
of this paper.  Anup Rao simplified an argument in the proof of
Lemma~\ref{lem:LocallyComputableCommonPartPre}.  Oded Regev and Ricky
Rosen found an error in a previous version of the proof of
Lemma~\ref{lem:ns_gameconditionedvalue}.  I was supported by the Swiss
National Science Foundation, project no.~200020-103847/1.

\appendix
\section{Non-triviality}
\label{app:nontriviality}
\paragraph{Local case}
We quickly reproduce a slight modification\footnote{Fortnow also lets
  the referee choose~$x=y=1$ with some probability, in which case the
  players cannot win the game.} of Fortnow's example \cite{Fortno89}
which shows that in general~$v(\mkG^2) > (v(\mkG))^2$.  The same
variation was also considered by Feige and Lov\'asz \cite{FeiLov92}.

The game we describe is over bits (i.e., all the queries and all the
responses are bits).  We
set~$\Pd_{XY}(0,0):=\Pd_{XY}(0,1):=\Pd_{XY}(1,0):=\frac13$, and define
\begin{align}Q(x,y,a,b) := \bigl((x \lor a) \neq (y \lor b)\bigr).
\end{align}
This can be described in words: Alice and Bob receive a bit, and at
least one of these bits is~$0$.  If both players receive~$0$,
exactly one player must respond with $1$.  If one of the players
receives~$1$, the other must respond with~$0$.  

We first show that for this game~$v=\frac{2}{3}$.  Clearly, $v \geq
\frac23$ (e.g.,~both players always answer~$0$).  To
show~$v\leq\frac23$ we check all deterministic strategies.  If both
players reply~$0$ on query~$0$, this fails in case~$x=y=0$ (and thus
with probability~$\frac13$).  If one player, w.l.o.g.~Alice,
answers~$0$ with~$1$ the players fail in case~$x=0$ and~$y=1$.

If this game is repeated twice in parallel, setting~$(a_1,a_2) :=
(x_2,x_1)$,~$(b_1,b_2) := (y_2,y_1)$ also wins with
probability~$\frac23$.  One can check this as follows: for every fixed
query~$(x_1,y_1)$ answering with~$(x_2,y_2)$ wins the first subgame
with probability~$\frac23$.  Moreover, with this strategy
$Q(x_1,y_1,a_1,b_1) \equiv
Q(x_2,y_2,a_2,b_2)$ which implies the claim.  

\paragraph{No-signaling case}
We now show that for the above game 
\begin{align}\label{eq:41}
  v(\mkG)= v(\mkG^2)=\vns(\mkG)=\vns(\mkG^2).
\end{align}
Previously, it was known that quantum strategies do not help Alice and
Bob to win this game \cite{Watrou02} (in both the single instance case
and where two parallel instances are used).  

To show~(\ref{eq:41}) it is sufficient to show that that $v(\mkG) =
\vns(\mkG)$ (since $\vns(\mkG^2)\leq \vns(\mkG)$ and $\vns(\mkG^2)
\geq v(\mkG^2) = v(\mkG)$ are already known).  There are two ways to
see that $v(\mkG) = \vns(\mkG)$.  First, one can notice that the joint
probability of Alice's and Bob's reply only matters if~$x=y=0$; i.e.,
only for one query.  In such a case one can always get a local
strategy which is as good as a given no-signaling strategy.
Alternatively, let $p$ be the probability that Alice replies~$0$ on
query~$0$ and $q$ be the probability that Bob replies~$0$ on
query~$0$.  In this case, the players win with probability at most~$p$
on query~$(x,y)=(0,1)$, with probability at most~$q$ on query~$(1,0)$,
and with probability at most $(1-p)+(1-q)$ on query~$(0,0)$, which
gives an overall winnig probability of at most $\frac23$.

\section{A Lemma on Relative Entropy}\label{app:relentrlemma}

This following lemma is well known, but we do not know of a standard
reference containing a proof of it.
\begin{lemma}
  Let $\Pd_{U^k} = \Pd_{U_1}\dots\Pd_{U_k}$ and
  $\Pd_{V^k}$ be distributions over the same set.  Then,
  \begin{align*}
    \sum_{j=1}^k D(\Pd_{V_j}\|\Pd_{U_j}) \leq
    D(\Pd_{V^k}\|\Pd_{U^k}).
  \end{align*}
\end{lemma}
\begin{proof}
  We prove the
  bipartite case; the general case follows by induction.
  \begin{align*}
    D(\Pd_{V_1 V_2}&\|\Pd_{U_1}\Pd_{U_2}) \\
    &=
    \sum_{(u_1,u_2)}
    \Pd_{V_1 V_2}(u_1,u_2)\log\Bigl(
    \frac{\Pd_{V_1 V_2}(u_1,u_2)}
    {\Pd_{U_1}(u_1)\Pd_{U_2}(u_2)}\Bigr)\displaybreak[3]\\
    &=
    \sum_{(u_1,u_2)}
    \Pd_{V_1 V_2}(u_1,u_2)\log\Bigl(
    \frac{\Pd_{V_1}(u_1)}
    {\Pd_{U_1}(u_1)}\Bigr)+
    \\ &\qquad\qquad+
    \sum_{(u_1,u_2)}\!\!
    \Pd_{V_1 V_2}(u_1,u_2)\log\Bigl(
    \frac{\Pd_{V_2|V_1=u_1}(u_2)}
    {\Pd_{U_2}(u_2)}\Bigr)\displaybreak[3]\\
    &=
    D(\Pd_{V_1}\|\Pd_{U_1})
    +
    \sum_{(u_1,u_2)}
    \Pd_{V_1 V_2}(u_1,u_2)\log\Bigl(
    \frac{\Pd_{V_2}(u_2)}{\Pd_{U_2}(u_2)}\cdot
    \frac{\Pd_{V_1 V_2}(u_1,u_2)}
    {\Pd_{V_1}(u_1)\Pd_{V_2}(u_2)}
    \Bigr)\\
    &=D(\Pd_{V_1}\|\Pd_{U_1})+D(\Pd_{V_2}\|\Pd_{U_2})
    +\!\!\sum_{(u_1,u_2)}\!\!
    \Pd_{V_1 V_2}(u_1,u_2)\log\Bigl(
    \frac{\Pd_{V_1 V_2}(u_1,u_2)}
    {\Pd_{V_1}(u_1)\Pd_{V_2}(u_2)}\Bigr)\\
    &\geq D(\Pd_{V_1}\|\Pd_{U_1})+D(\Pd_{V_2}\|\Pd_{U_2}),
  \end{align*}
  where the last inequality follows from the log-sum inequality (see
  \cite[Theorem 2.7.1]{CovTho91}).
\end{proof}

\section{Converse of Lemma~\ref{lem:fractionalCovEq}}
\label{app:fractEquivalence}

In this appendix we show that Lemma~\ref{lem:fractionalCovEq} can be
strengthened to get an ``if and only if'' condition.
\begin{lemma}\label{lem:converse}
  Let a conditional distribution $\Pd_{Z|AB}$ be given.  If, for all
  product distributions~$\Pd_{AB}=\Pd_{A}\Pd_{B}$ the Markov condition
  $A\leftrightarrow Z \leftrightarrow B$ is satisfied, then  there
  exists functions $f(a,z): \cA\times\cZ \rightarrow [0,1]$
  and~$g(b,z): \cB\times\cZ\rightarrow[0,1]$ such that
  \begin{align}
    \Pd_{Z|A=a B=b} = f(a,z)\cdot g(b,z).
  \end{align}
\end{lemma}
\begin{proof}
  Fix an arbitrary~$z$ throughout the proof, and consider
  arbitrary elements $a,a'\in\cA$ and $b,b' \in \cB$.  We
  set~$\Pd_{A}(a) = \Pd_{A}(a') = \frac12$ and $\Pd_{B}(b) =
  \Pd_{B}(b') = \frac12$.  The Markov condition implies
  \begin{align*}
    \Pd_{A|Z=z B=b}(a) = \Pd_{A|Z=z B=b'}(a)
  \end{align*}
  which is equivalent to 
  \begin{align*}
    \frac{\Pd_{ABZ}(a,b,z)}{\Pd_{ABZ}(a,b,z)+\Pd_{ABZ}(a',b,z)} =
    \frac{\Pd_{ABZ}(a,b',z)}{\Pd_{ABZ}(a,b',z)+\Pd_{ABZ}(a',b',z)}
  \end{align*}
  or (because of our choice of~$\Pd_{AB}$)
  \begin{align*}
    \frac{\Pd_{Z|A=a,B=b}(z)}{\Pd_{Z|A=a,B=b}(z)+\Pd_{Z|A=a',B=b}(z)} =
    \frac{\Pd_{Z|A=a,B=b'}(z)}{\Pd_{Z|A=a,B=b'}(z)+\Pd_{Z|A=a',B=b'}(z)}.
  \end{align*}
  Analogously one gets (by swapping the roles of~$a$ and~$a'$)
  \begin{align*}
    \frac{\Pd_{Z|A=a',B=b}(z)}{\Pd_{Z|A=a,B=b}(z)+\Pd_{Z|A=a',B=b}(z)} =
    \frac{\Pd_{Z|A=a',B=b'}(z)}{\Pd_{Z|A=a,B=b'}(z)+\Pd_{Z|A=a',B=b'}(z)}.
  \end{align*}
  Together, this implies
  \begin{align}\label{eq:16}
    \Pd_{Z|A=a,B=b}(z) \Pd_{Z|A=a',B=b'}(z) = 
    \Pd_{Z|A=a,B=b'}(z) \Pd_{Z|A=a',B=b}(z).
  \end{align}
  
  Fix now~$z$ and let~$f(\cdot,z)$ and~$g(\cdot,z)$ be functions onto~$[0,1]$
  which satisfy
  \begin{align}\label{eq:40}
    f(a,z)g(b,z) \geq \Pd_{Z|A=a,B=b}(z)
  \end{align}
  for all~$(a,b)$ and for which the number of pairs $(a,b)$ for which
  $f(a,z)g(b,z) > \Pd_{Z|A=aB=b}(z)$ is minimal (such functions exist
  since $f(a,z) = g(b,z) = 1$ satisfy~(\ref{eq:40})).  We assume this
  number is non-zero and obtain a contradiction.  For this,
  let~$(a_1,b_1)$ be a pair for which $f(a_1,z)g(b_1,z) > 0$ and for
  which the quotient $\Pd_{Z|A=a_1,B=b_1}(z)/f(a_1,z)g(b_1,z) < 1$ is
  minimal.

  We define
  \begin{align*}
    f'(a,z) &:= \begin{cases}
      f(a,z) & \text{if $a\neq a_1$,}\\
      \frac{\Pd_{Z|A=a,B=b_1}(z)}{g(b_1,z)} & \text{if $a = a_1$}
    \end{cases}\\
    \intertext{and}
    g'(b,z) &:= \begin{cases}
      g(b,z) & \text{if $b\neq b_1$,}\\
      \frac{\Pd_{Z|A=a_1,B=b}(z)}{f(a_1,z)} & \text{if $b=b_1$}.
    \end{cases}
  \end{align*}
  We note that $f'$ and~$g'$ cannot take values larger than~$1$.  For
  example, $f'(a_1,z)>1$ implies $\Pd_{Z|A=a_1,B=b_1}(z)>g(b_1,z) \geq
  f(a_1,z) g(b_1,z)$, which contradicts~(\ref{eq:40}).  We further
  claim that either the pair $(f',g)$ or $(f,g')$ still
  satisfies~(\ref{eq:40}).  Otherwise, there are values~$a_2$
  and~$b_2$ such that
  \begin{align*}
    f'(a_1,z)g(b_2,z) &= \frac{\Pd_{Z|A=a_1,B=b_1}(z)}{g(b_1,z)}
    g(b_2,z) > \Pd_{Z|A=a_1 B=b_2}(z) \\
    \intertext{and}
    f(a_2,z)g'(b_1,z) &= f(a_2,z) \frac{\Pd_{Z|A=a_1,B=b_1}(z)}{f(a_1,z)}
     > \Pd_{Z|A=a_2 B=b_1}(z),
  \end{align*}
  which implies
  \begin{align*}
    \frac{\Pd_{Z|A=a_1,B=b_1}(z)}{g(b_1,z)}\frac{\Pd_{Z|A=a_1,B=b_1}(z)}{f(a_1,z)}
        g(b_2,z)f(a_2,z) > \Pd_{Z|A=a_2 B=b_1}(z) \Pd_{Z|A=a_1 B=b_2}(z),
  \end{align*}
  and using~(\ref{eq:16}) 
  \begin{align*}
    \frac{\Pd_{Z|A=a_1,B=b_1}(z)}{f(a_1,z)g(b_1,z)}
     >\frac{\Pd_{Z|A=a_2,B=b_2}(z)}{g(b_2,z)f(a_2,z)}
  \end{align*}
  contradicting the way we chose $(a_1,b_1)$.  Thus, either $(f',g)$
  or~$(f,g')$ still satisfies (\ref{eq:40}) and since the
  respective version of (\ref{eq:40}) is satisfied with equality for
  at least one more pair $(a,b)$ (namely for $(a_1,b_1)$) than for
  which $(f,g)$ satisfies it, we get a contradiction.
\end{proof}
\end{document}